%% file: Manuscript.tex
\documentclass[12pt]{article}
\usepackage[a4paper, total={6.3in, 9.7in}]{geometry}
\usepackage[utf8]{inputenc}
\usepackage[english]{babel}
\usepackage{blindtext}
\usepackage{amsmath}
\usepackage{physics}
\usepackage{chemformula}

\usepackage{xcolor}
\usepackage{soul}
\usepackage[T1]{fontenc} 
\usepackage{graphicx}
\usepackage{authblk}
\usepackage[backend=biber, sorting=none, style=nature]{biblatex}
\usepackage[compact]{titlesec}
\usepackage{float}

\usepackage[font=footnotesize,labelfont=bf]{caption}

\usepackage[hypertexnames=false,colorlinks=true,allcolors=blue]{hyperref}
\DeclareUnicodeCharacter{0301}{************}

\newcommand{\beginsupplement}{%
	\setcounter{table}{0}
	\renewcommand{\thetable}{S\arabic{table}}%
	\setcounter{figure}{0}
	\renewcommand{\thefigure}{S\arabic{figure}}%
	\setcounter{section}{0}
	\renewcommand{\thesection}{S\arabic{section}}
	\setcounter{equation}{0}
	\renewcommand{\theequation}{S\arabic{equation}}
}

\title{Anisotropic transport and Negative Resistance in a polycrystalline metal - semiconductor (Ni-TiO$_2$) hybrid}

\author[1,4]{Harikrishnan G}

\author[1]{Shashwata Chattopadhyay}
\author[2]{K. Bandopadhyay}
\author[2]{K. Kolodziejak} 
\author[2.3]{Dorota A. Pawlak}
\author[1,5]{J. Mitra}

\affil[1]{%
	\textit{School of Physics, Indian Institute of Science Education and Research, Thiruvananthapuram, Kerala 695551, India}}
\affil[2]{%
	\textit{Institute of Microelectronics and Photonics, University of Warsaw, Poland}}
\affil[3]{%
	\textit{$Ensemble^3$ Centre of Excellence, Warsaw, Poland}}
\affil[4]{harikrishnan17@iisertvm.ac.in}
\affil[5]{j.mitra@iisertvm.ac.in}

\date{}
\begin{document}
	\maketitle
	\begin{abstract}
		We investigate anomalous electrical transport properties of a Ni – \ch{TiO2} hybrid system displaying a unique nanostructured morphology. The system undergoes an insulator to metal transition below 150 K with a low temperature metallic phase that shows  negative resistance in a four-probe configuration. Temperature dependent transport measurements and numerical modelling show that the anomalies originate from the dendritic architecture of the \ch{TiO2} backbone interspersed with Ni nanoparticles that paradoxically renders this polycrystalline, heterogeneous system highly anisotropic. The study critiques inferences that may be drawn from four-probe transport measurements and offers valuable insights into modelling conductivity of anisotropic hybrid materials.
	\end{abstract}
	\begin{center}
		\textbf{Keywords:} Anisotropy, Hybrid material, Insulator-Metal Transition, Absolute Negative Resistance
	\end{center} 
	\input{Main}
	\pagebreak
	\beginsupplement
	\input{SI}
	\printbibliography
\end{document}

%% file: Main.tex
Inhomogeneous media are multicomponent systems whose composition, structure and thus properties vary spatially across length scales. They present a complex domain of hybrid material systems where the response of the system is often more than the sum of its parts. For example, the electrical transport properties of inhomogeneous systems are not straightforward to comprehend in terms of a Drude like model where electrical conductivity ($\sigma_E=nq\mu$) is estimated in terms of carrier density ($n$) and mobility ($\mu$)  since both variables may be spatially dependent. However, ascribing an effective $\sigma_E$, that averages over response of individual components and their volume fractions offers a quantifiable and useful parameter to describe the macroscopic conductivity. Various analytical and numerical schemes like effective medium theories, resistor network models, Monte Carlo simulations and finite element or finite difference methods have been used to estimate $\sigma_E$ and gain insights into transport properties \cite{palla2016transport,kim2020anomalous,zhang2018shell,ahmadi2020stochastic,montes2018electrical,korchagin2021mathematical}. Similarly, thermal transport is impacted due to the presence of junctions and interfaces and the shorter order parameters that scatter electrons and phonons differentially. However, these complexities offer additional opportunities to tailor and control electric and thermal transport differentially and has lead to improved performance of thermovoltaic and photo-thermovoltaic devices \cite{zhang2017manipulation,silk2008thermoelectric,han2017thermoelectric,harada2010thermoelectric,omari2011photothermovoltaic}. Understanding physical properties of inhomogeneous materials holds immense significance across a wide spectrum of applications, ranging from electronic and energy devices to photonic,  optoelectronic and optomechanical systems. 
Hybrid materials have been synthesized in complex forms like alloys, high-entropy alloys, eutectics, composites, twisted fibres and 3d nanostructures  \cite{morales2005hydrogen,yao2021extreme,gibson2010review,zhang2018shell,kim2020anomalous,kolodziejak2016synthesis}, employing both bottom-up mixing and top-down processes. Ubiquitous to these complex systems are the interfaces at which dissimilar atoms, clusters or atomic planes meet, which crucially control their overall property including their electrical identity.

This investigation reports on the electrical properties of a bulk nanostructured hybrid of Magn$\acute{e}$li phase titanium dioxide (\ch{mTiO2}) background  decorated with Ni nanoparticles (NP). Surprisingly, this heterogeneous and polycrystalline sample exhibits anisotropic electrical transport accompanied by a insulator-metal transition (IMT) with a low T metallic phase and negative resistance therein. 
The samples were prepared from a phase segregated eutectic of \ch{NiTiO3 - TiO2}, synthesized employing the micro-pulling-down ($\mu$-PD) method \cite{yoon1994crystal,kolodziejak2016synthesis}, by reducing  in \ch{H2} environment at 1050$^\circ$C\cite{arvanitidis2000intrinsic,morales2005hydrogen}.
 
The reduced sample has a porous bulk composed of fibrous strands of \ch{mTiO2} decorated with polydispersed Ni NPs, a distributed Schottky junction system. The intrinsic transport anisotropy is investigated via temperature dependent transport measurements in varied geometry. 
Associated modelling of electrical transport and reconstruction of the temperature dependent conductivity tensor reveals that the negative resistance arises from the anisotropy that increases in the low temperature regime. The anisotropy is surprising given the heterogeneous and polycrystalline nature of the sample and the observed negative slope of the voltage - current $VI$ characteristics question the tacit assumptions of the conventional four probe resistance and transport measurement schemes. 
The low T metallic phase is shown to be triggered by a changeover in the thermal expansion coefficient ($\alpha$) of the sample that becomes negative close to the IMT transition temperature (T$_c\sim$ 100 K). These features stem from the dendritic porous architecture of the sample with the polydispersed Ni NPs and the resulting variable transport pathways. The material morphology also induces pressure sensitivity to the sample resistance $\sim$ 0.096 kPa$^{-1}$, that is comparable to the sensitivity of similar disordered systems \cite{liu2017ultrafast}. 
Engineering porosity and interfaces is also an emerging area in thermo-voltaic materials design \cite{zhang2017manipulation} since the heterogeneous architecture enables high thermal gradient while maintaining charge transport. Such multi functional hybrids  are key to devices with applications ranging from sensors, energy generation, storage\cite{saveleva2019hierarchy,gu2018introducing,nicole2014hybrid,gibson2010review}. etc.

The eutectic samples were cut into circular disks of diameter $\sim$ 3 mm and thickness $\sim$ 1 mm, polished and reduced thereafter (see Supplementary Information (SI) section S1 for sample details). 

\begin{figure}
\begin{center}
	\includegraphics[width=\textwidth]{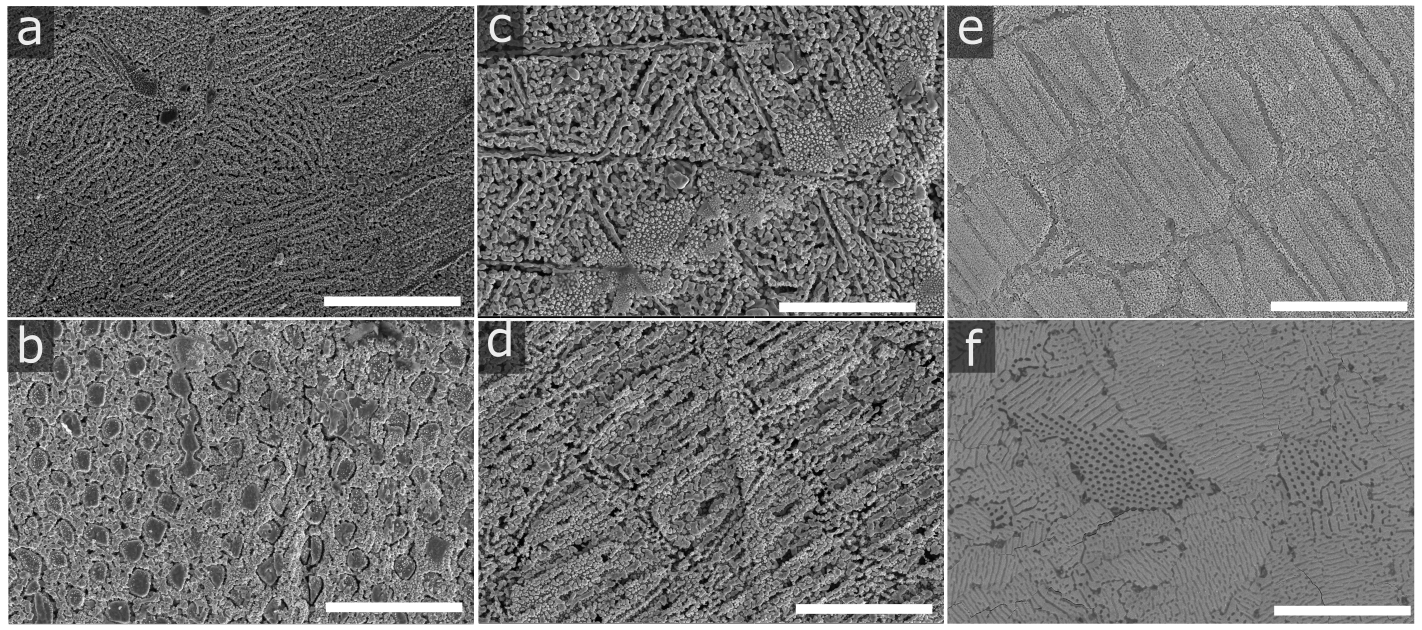}
	\caption{Secondary electron images showing variation amongst \ch{H2} annealed NTT  samples (scale 30$\mu$m).}
	\label{fig:1}
\end{center}
\end{figure}
The secondary electron micrographs in fig. \ref{fig:1} show the varied morphology across samples prepared under different conditions as detailed in SI table S1. SI fig. S2 shows the higher resolution images corresponding  to fig. \ref{fig:1}, with globular Ni nanostructures decorating different \ch{mTiO2} morphologies. The electrical properties of samples annealed at 1050 $^{\circ}$C (fig. \ref{fig:1}(a,c,d)) designated as NTT1050, which all show fibrous \ch{mTiO2} structures are discussed. As seen in figs. \ref{fig:1}(a,c,d) the fibrous \ch{mTiO2} structures possess short range order ($\sim 30 \mu$m) but change orientation at longer length scales, developing a  dendritic network  across the sample. In each case the Ni NPs vary in diameter from 50 nm to $\sim 1 \mu$m as shown in the atomic force microscopy (AFM) topography images in SI fig. S3(a-c). 
The polycrystalline x-ray diffraction spectra of the as prepared eutectic sample and those annealed at 700 $^\circ$C and 1050 $^\circ$C in SI fig. S5(a) show the evolution of the initial phase separated \ch{NiTiO3} - rutile \ch{TiO2} system reduced to \ch{mTiO2} - Ni NPs in the final form. Variation in crystal structure across the final \ch{mTiO2}-Ni samples, with the multiple Magn$\acute{e}$li and sub-oxide phases in \ch{mTiO2} are shown in SI fig. S5(b).      
The energy dispersive x-ray spectrum and elemental map (SI fig. 4) on NTT1050 delineates the chemical identity of the NPs from the background matrix and quantify the relative abundance of Ti:O:Ni $\sim$ 1.0:2.0:1.3. 
Conducting AFM current maps, shown in SI fig. S3(d) evidence the Ni NPs  as locally higher conductance regions in lower conducting background. The current maps are conspicuous by the absence of any  filamental high conductance networks across the surface indicating that the transport remains a bulk phenomena in these samples.
X-ray photoelectron spectroscopy data (SI fig. S6) evidence Ti  primarily in the Ti$^{4+}$ state, with the weaker Ti$^{3+}$ and Ti$^{2+}$ signatures originating from the mixed stoichiometry of the various \ch{mTiO2} phases. A majority of Ni appears in the metallic Ni$^0$ state with a fraction as Ni$^{2+}$ \cite{pan1992interaction,biesinger2009x,biesinger2010resolving}. 
The results together attest to the porous morphology of the samples with the fibrous, dendritic architecture of the \ch{mTiO2} background and polydispersed Ni NPs. These structures extend through the bulk of the samples  as seen in the SE images in SI fig. S7 along broken edges.  

It is worth noting that as seen in the XRD plots (SI fig. S5b), there is only statistical similarity in distribution of the various Magn$\acute{e}$li and sub-oxide phases of \ch{TiO2}, and Ni NPs across the various samples investigated here, which are polycrystalline and inhomogeneous in nature.

\begin{figure}
\begin{center}
	\includegraphics[width=0.98\textwidth]{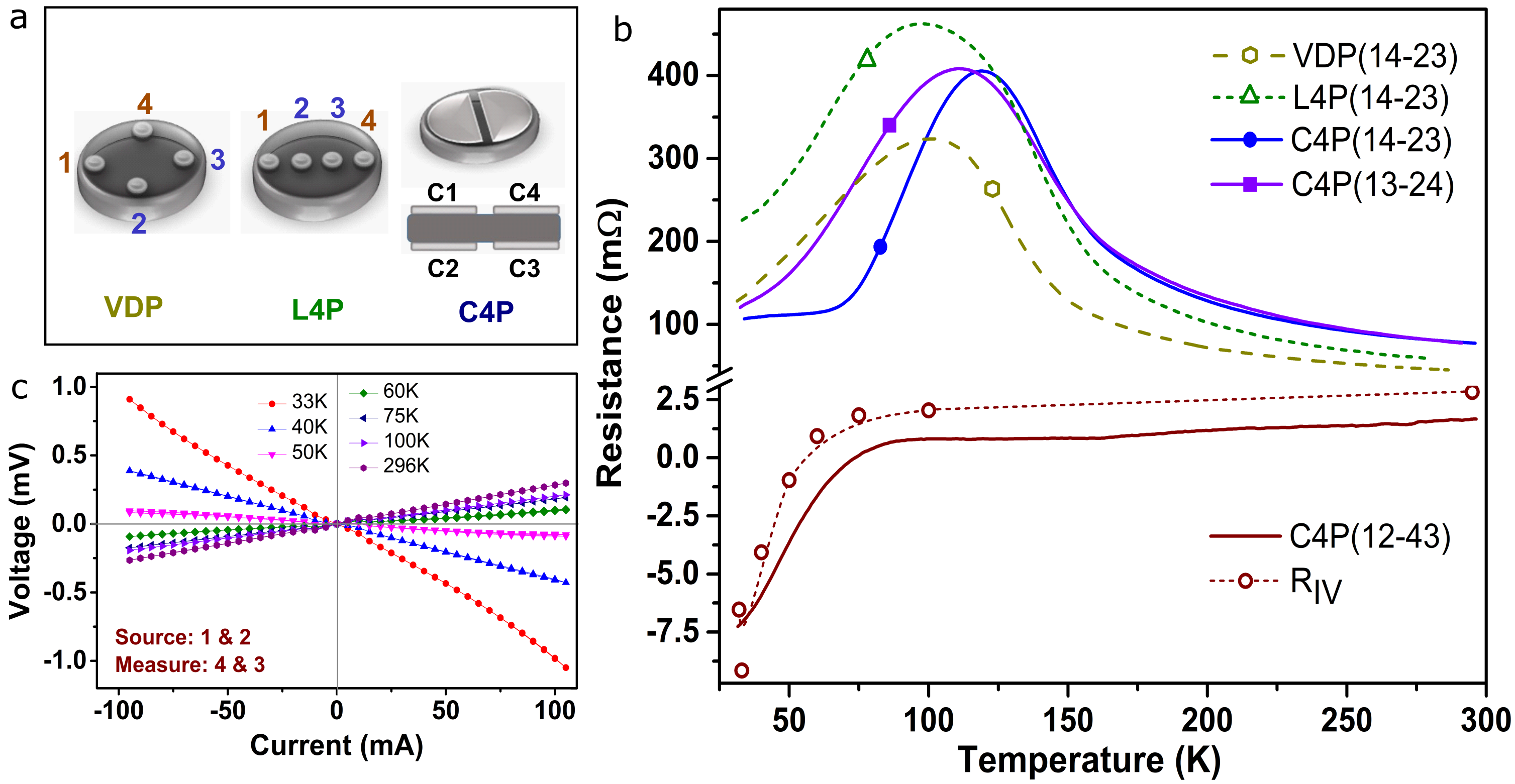}
	\caption{(a) 4 probe electrical contact configurations on NTT1050. (b) Corresponding R \textit{vs} T plots and slopes from (c) which shows the T stabilized $VI$s (R$_{IV}$) of C4P (12-43).}
	\label{fig:2}
\end{center}
\end{figure}
The NTT1050 samples are $n$-type semiconductors with carrier density $\sim 10^{-19}$/cc with two and four probe resistances $\sim$ 1 $\Omega$ and $\sim$ 100 m$\Omega$, respectively. Details about electronic and optoelectronic properties of NTT1050 are available in a recent communication in \textit{arXiv} \cite{g2023anomalous}. Fig. \ref{fig:2}(a) show three different contact configurations in which the transport properties were investigated, with the corresponding resistance vs. temperature ($RT$) plots shown in fig. \ref{fig:2}b. 
All $RT$ plots were recorded with 10 mA current and the legends in the figure identify the contact geometry with the four numerals denoting the contact pad used to source and sink current and measure voltage as (I$_+$,I$_-$ -- V$_+$,V$_-$). The Van der Pauw (VDP) and linear four probe (L4P) configurations  evidence insulating (semiconducting) behaviour (dR/dT $<$ 0) from 300 K down to $\sim$ 100K, and below a transition temperature (T$_c$) dR/dT becomes positive indicating a temperature driven insulator to metal transition (IMT). 

Response in the high T regime is likely dominated by transport through the \ch{mTiO2} network and the conductivity attributed to that of the defect laden black \ch{TiO2} (sub-oxides, magneli) phases in the sample\cite{xu2016structures,rajaraman2020black,lu2016conducting}, since oxygen deficiency is known to increase carrier density in metal oxide systems, even affecting their temperature dependencies\cite{harada2010thermoelectric,he2007thermoelectric}.  
Previous reports of IMT in \ch{mTiO2} systems, especially \ch{Ti2O3} have recorded significantly higher transition temperatures (T$_c$) than that observed in these samples \cite{mott1981metal, uchida2008charge, chang2018c, shvets2020suppression}. Its also worth noting that bare \ch{TiO2} and the as prepared eutectic \ch{NiTiO3} - \ch{TiO2} samples show insulating behaviour down to T $\sim$ 20 K (SI fig. S8a) and the NTT1050 sample shows negligible magnetoresistance as shown in SI fig. S8b, negating a magnetic interaction driven transition. The results also suggest that appearance of the metallic phase is linked to the porous dendritic architecture as much with the intrinsic material properties of \ch{mTiO2} and Ni NPs.

RT measured across the sample in the C4P geometry  
in the (14 - 23) and (13 - 24) configurations follow those from the planar measurements in the VDP and L4P configurations as shown in fig. \ref{fig:2}b. However, the resistance measured in the (12 - 43) configuration, at room temperature, shows R$_{1243} \sim 1 m\Omega$ which is lower than R$_{1423}$ by a factor of 100, indicating that the conductivity is higher along the perpendicular to the plane of the disk than along the plane of the disk, and consequently the sample exhibits anisotropic conductivity. 
Unlike the in-plane measurements, the RT for R$_{1243}$ shows metallic behaviour in the  high temperature regime with dR/dT $>$ 0.  Further, R$_{1243}$ becomes negative below T $\sim$ 75 K.  And a similar trend is observed for resistance calculated from the slope of the $VI$ characteristics measured at different T, as co-plotted in the same fig. \ref{fig:2}b. 
The $VI$s obtained by sourcing current between contacts C1 and C2 and voltage measured between C4 and C3 are plotted in fig. \ref{fig:2}c, which clearly show a decrease in slope with decreasing T, with the slope being turning negative for the $VI$ at 50 K and below. The $VI$ plots confirm that the negative values of R$_{1243}$ at low T do not arise due to measurement errors such as device offsets etc. 
SI fig. S9(a-b) shows the $VI$ characteristics at 300 K and 33 K measured in (14-23) and (13-24) configurations which do not exhibit negative slope (negative resistance) at low temperatures i.e. in the metallic phase of the corresponding $RT$ plots of fig. \ref{fig:2}b.

\begin{figure}
\begin{center}
	\includegraphics[width=0.8\textwidth]{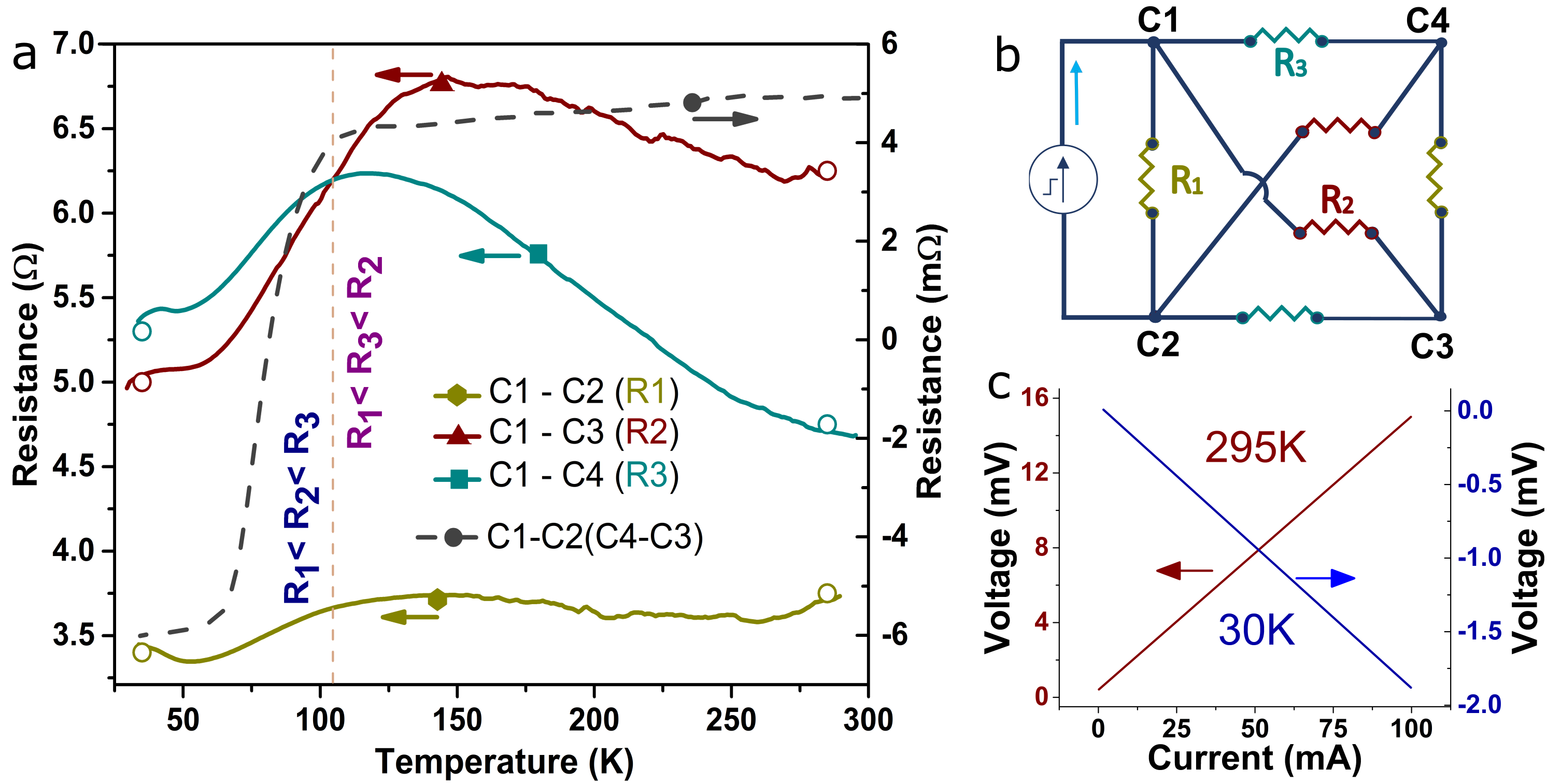}
	\caption{(a) 2 probe(left axis) and 4 probe (right axis) R\textit{vs}T plots. (b) The resistor network model (RNM) and (c) the Multisim circuit simulated 4 probe $VI$ with experimentally obtained R1, R2 \& R3 values at 35K and 295K.}
	\label{fig:3}
\end{center}
\end{figure}
To further probe the anisotropy and the negative resistance quasi four probe measurements were conducted in the C4P configuration where current source and voltage measurements were conducted using 2 contacts. Fig. \ref{fig:3}a shows the RT plots for resistances measured across  in-plane contacts C1 - C4 (R3), across the vertical contacts C1 - C2 (R1) and across the diagonal contacts C1 - C3 (R2).
Though R1, R2, R3 have comparable values, the R2 and R3 variation have dR/dT $<$ 0 near room temperature in contrast to R1 that shows dR/dT $>$ 0. At lower T  both R2 and R3 reproduce the IMT, while R1 shows low temperature dependence down to 20 K. Fig. \ref{fig:3}a also co-plots the four probe RT for R$_{1243}$ (right axis) showing a small positive dR/dT $\sim$ 4E$^{-6}$ $\Omega$/K (metallic transport) at high T, which  increases to dR/dT $\sim$ 2E$^{-4}$ $\Omega$/K at T$_1$ close to the MIT temperature, and finally R$_{1243}$ $<$ 0 below 70 K. Note that for T $<$ T$_1$, R1 $<$ R2 $<$ R3 and R1 $<$ R3 $<$ R2 above T$_1$, and this change in the relative values of R2 and R3 is intrinsic to the anisotropy and negative resistance regime at lower temperatures. The difference in T dependence of resistance along the in-plane and vertical direction of the sample  is also evidenced in four probe measurements as shown in SI fig. S10 across multiple samples investigated here. While in-plane measurement (R$_{1423}$) show insulating behaviour between 300 K - 100 K, followed by MIT at lower T (SI fig. S10a), vertical measurement (R$_{1243}$) show metallic transport between  300 K - 100 K  (SI fig. S10b). This difference in the intrinsic nature of transport along the in-plane and vertical direction in two and four probe measurements establishes the inherent conductance anisotropy of these samples.   
$VI$ in the C4P configuration provides additional insights to the anisotropy in transport properties i.e. conductivity. 
SI fig. S11 shows a series of $VI$ for source - sense configurations (14-23), (12-43), (13-24) and (13-42) at 300 K.  Both (14-23) and (12-43) data show $VI$s with positive slope with R$_{1423}$ $>$ R$_{1243}$ again confirming higher vertical conductance compared to in-plane value by a factor of $\sim$ 100. With  contact C1 and C3 as the current source and sink, the voltage developed is higher at C2 than at C4 and thus R$_{1324}$ is positive while R$_{1342}$ has a negative value. The corresponding schematic in SI fig. S11 labelled 1324 shows the current pathway, where current injected into the sample at C1 travels vertically more than horizontally commensurate with the anisotropic conductivity and creates voltage distribution across the sample.   
To comprehend the link between anisotropic conductivity and negative resistance (R$_{1243}$) in the low temperature metallic phase we resort to a single cell resistor network model (RNM), shown in fig. \ref{fig:3}b which mimics a vertical cross section through the sample. Here, the resistors R1, R2 and R3 refer to the vertical, diagonal and horizontal resistances as depicted in the $RT$ plot in fig. \ref{fig:3}a. As indicated on the schematic points (contacts) C1 and C2 are used to drive current and C4 and C3 used to sense voltage. The values of the resistors chosen at 295 K and 35 K are motivated by the $RT$ and indicated by open circles in fig. \ref{fig:3}a, and tabulated in SI table S2. The ensuing $VI$ characteristics (fig. \ref{fig:3}c) evidence a positive slope at 295 K and negative slope at 35 K, the later induced by the swap in relative magnitude of R2 and R3 with R1 remaining unchanged.  Further details of the RNM analysis is available at SI section S5.
Note that the two diagonal resistors (R2) in the RNM bypass each other and do not connect at a node at the centre of the cell and the negative slope is not realised if they interconnect to a node. 
Thus the R2 resistors provide the necessary alternate pathways connecting points C1 - C3 and C2 - C4, in parallel to the R1, R3 combination.   
Such alternate transport pathways are not physical in a 2d system but is realizable once we include transport along the third dimension.

\begin{figure}
\begin{center}
	\includegraphics[width=0.78\textwidth]{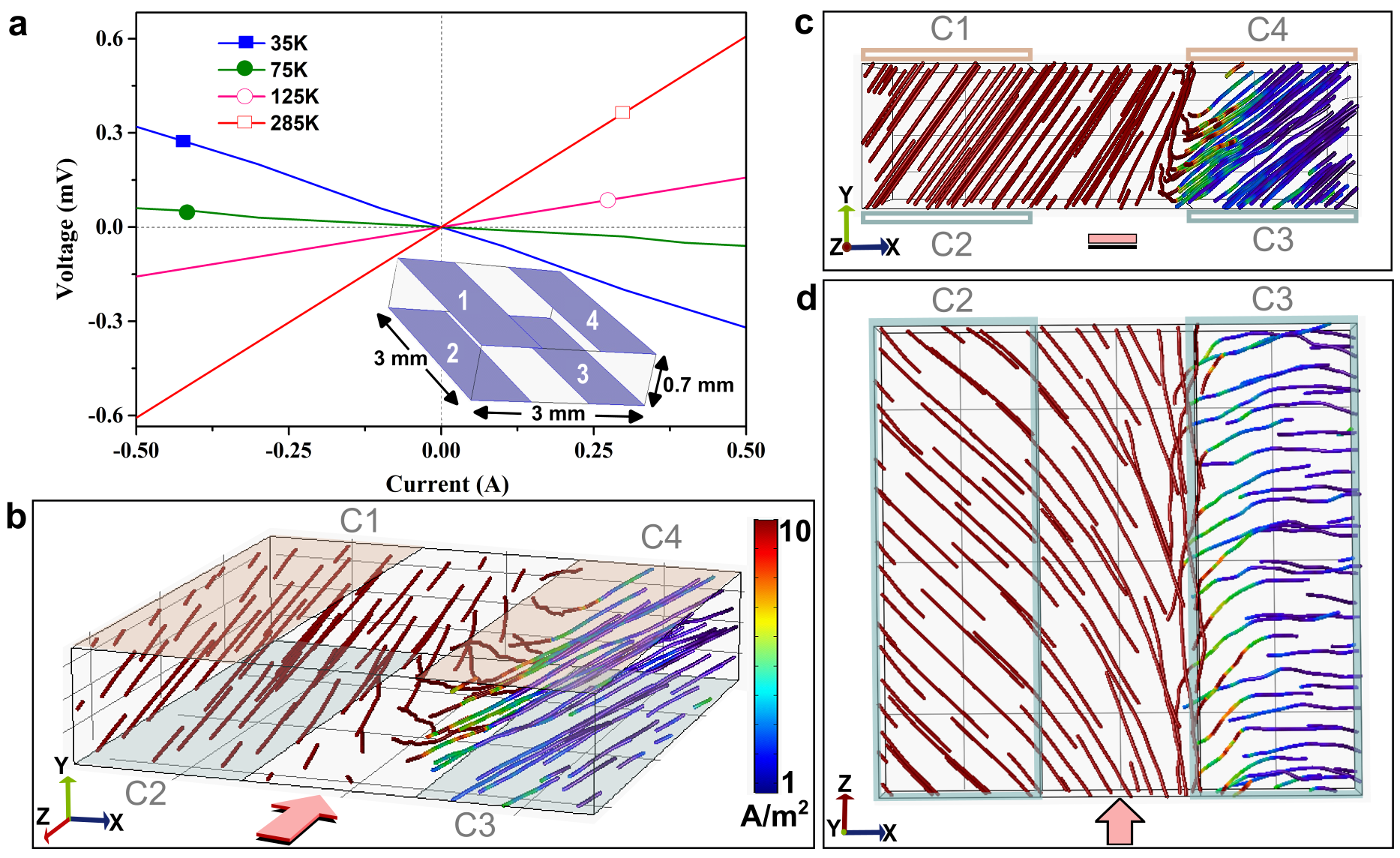}
	\caption{a: Simulated voltage - current characteristics at various temperatures with corresponding $\sigma$(T), (b-d) Simulated 3d plots showing current streamlines in the sample at 35 K, in different orientations to visualize the anisotropic electrical transport. C1 - C4 denote the contact pads in the C4P configuration.}
	\label{fig:4}
\end{center}
\end{figure}

Transport across a 3d anisotropic model with the sample envisaged as a continuous rectangular slab (inset of fig. \ref{fig:4}a) was developed with the transport properties simulated using finite element method (FEM) analysis using  COMSOL Multiphysics\textsuperscript{\textregistered} software. 
The dimensions of the slab ($x,y,z$) are 3 mm $\times$ 0.7 mm $\times$ 3 mm, mimicking the physical dimensions of the sample along with four extended contacts C1 - C4 in the C4P configuration. The anisotropic conductivity tensor ($\sigma$) is defined as a 3 $\times$ 3 non-diagonal, symmetric matrix with six independent elements. It is non-trivial to solve the inverse problem of re-constructing unique $\sigma$(T) matrices that will reproduce the observed transport anisotropy (fig. \ref{fig:2} and fig. \ref{fig:3}) at various T for this system. Using a recursive elimination method, $\sigma$ at four temperatures were obtained that preserve the relative magnitudes of the principal conductivity elements and reproduce the primary transport features i.e. negative resistance in the (12-43) configuration. Simulated $VI$s with $\sigma$ obtained at 35 K, 75 K, 125 K and 285 K are shown in fig. \ref{fig:4}a with the corresponding $\sigma$ tabulated in SI table S3. For each value of the current sourced between C1 (I$_+$) and C2 (I$_-$), the voltage values of the $VI$ plots are calculated as the difference between the electric potential integrated over the area of contacts C3 and C4, taken as equivalent to the measured voltage. 
It has been confirmed that all $\sigma$(T) matrices are diagonalizable as $\sigma_P$(T), with the principal elements having real and positive values. Further, the associated transformation matrices satisfy all properties to be Eulerian rotation matrices, as discussed in SI section S5.3. 
An anisotropic single crystal cut in an orientation not matching the principal axes would evidence a symmetric, non-diagonal $\sigma$(T), which is observed here for a polycrystalline and inhomogeneous system. The anisotropic electrical transport through the sample at 35 K are shown in figs. \ref{fig:4}(b-d) that plot the simulated current streamlines in various orientations. A majority of the current flows directly from C1 to C2 owing to the higher effective conductance along $y$ direction compared to $x$ direction. The higher conductance along $z$ is evidenced by the tilt in the streamlines in the $xz$ plane plot fig. \ref{fig:4}d. The streamlines assume a curved topology away from the contacts C1, C2 and  bends towards C3 with higher magnitude than towards C4. For the same measurement configuration, SI fig. 19 shows the simulated current streamlines at 285 K, for the corresponding $\sigma$ matrix, depicting a more conventional yet anisotropic transport with the $VI$ shown in fig. \ref{fig:4}a with a positive slope. SI fig. S18 show the resulting voltage distribution along a $xy$ cross section of the sample at all four temperatures. For identical current injection,  while $\Delta V$ = V$_4$ - V$_3$ is positive for the $\sigma$ corresponding to T = 285 K and 125 K, it is negative for T = 75 K and 35 K. The voltage and current streamline plots for 35 K  also predict that for the C4P configuration if the in-plane contact pairs (C1, C4) and (C2, C3) are in close proximity, separation $<$ 1 mm the negative value of V$_4$ - V$_3$ may turn positive, the limiting case being the quasi-four probe measurement across two contacts C1 - C2 (R1) as shown in fig. \ref{fig:3}a. SI fig. S20 shows the $RT$ for such a configuration where the R$_{1243}$ remains positive at all temperatures. Thus the  inhomogeneous, dendritic  nature of the porous sample with a non-trivial multiply-connected topology is responsible for the observed anisotropy and negative resistance. These observations and analysis are similar to a recent report of negative resistance in bundles of twisted niobium nitride - carbon nanotube yarns \cite{kim2020anomalous}.

Previous reports on metal-insulator transition in titanium oxides, especially titanium sesquioxide (\ch{Ti2O3}) \cite{mott1981metal,chang2018c} all record a metallic phase at high T with a  low T insulating phase with the transition being variously associated with crystallographic change \cite{uchida2008charge,shvets2020suppression} and antiferromagnetic ordering \cite{morin1959oxides}. In the present case, the $RT$ in fig. \ref{fig:2}b shows a broad transition with a low T metallic phase contrary to the above and a T$_c$ (T at which dR/dT = 0)  that varies with the transport measurement direction by 25 K. Further, the highly porous architecture of the sample with polydispersed Ni NPs indicate that the sample may be deformable under temperature or pressure induced strain. SI fig. S13 shows T dependent thermal expansion ($\Delta L/L$) and the thermal expansion coefficient ($\alpha$) for the sample from 250 K - 50 K. While $\alpha$(T) is positive near 250 K, it goes to zero around 125 K and becomes negative at lower T, which indicates that the sample expands on cooling below 125 K. 
\begin{figure}
\begin{center}
	\includegraphics[width=0.99\textwidth]{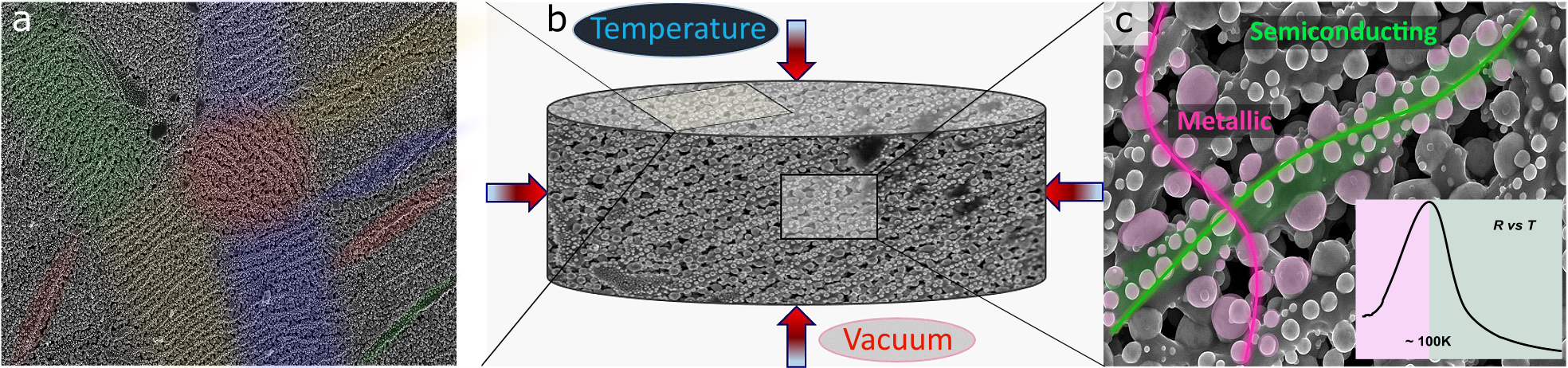}
	\caption{(a) False colored SEM image differentiating local order \ch{mTiO2} fibres across a sample, (b) cartoon of a sample reconstructed using SEM images to show the porous, dendritic architecture, (c) false coloured SEM image showing different transport pathways, inset plots a R vs. T showing insulator metal transition.}
	\label{fig:5}
\end{center}
\end{figure}
Based on the SEM images shown in fig. \ref{fig:5} and transport data it is evident that the fibrous network of \ch{mTiO2} form the primary transport pathway around room temperature.  As the sample is cooled the sample along with the \ch{mTiO2} fibre network contract and transport is dominated by that through the semiconducting matrix. Cooling below T $\simeq$ 125 K would now expand the previously contracted \ch{mTiO2} fibre network thus decreasing sample porosity. It is hypothesized that this expansion brings the Ni NPs, around the dendritic fibres, in closer proximity to form new percolating transport pathways. As depicted in fig. \ref{fig:5}c these additional conduction channels lower sample resistance with decreasing temperature (dR/dT $>$ 0) and form the dominant transport pathway in the  low T regime. Thus the metallic state is a manifestation of transport through the disordered metal channel of percolating Ni NPs.

\begin{figure}
\begin{center}
	\includegraphics[width=0.7\textwidth]{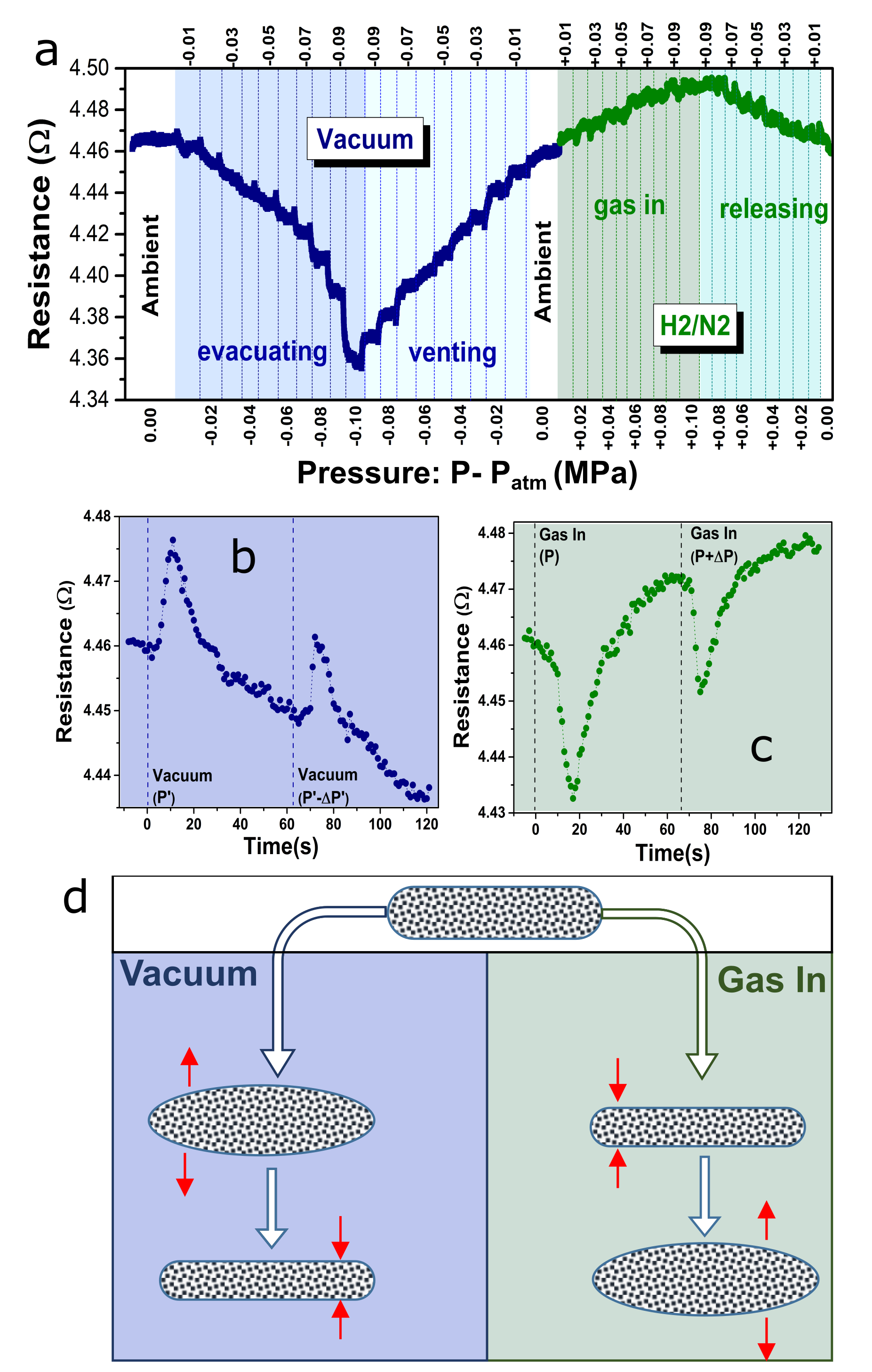}
	\caption{(a) Plot of R \textit{vs} P showing the effect of pressure (vacuum). Separate analysis of R \textit{vs} time with (b) vacuum and (c) gas. (d) Visualization of probable mechanism behind observed effects.}
	\label{fig:6}
\end{center}
\end{figure}
Evolution of sample resistance with pressure adds credence to the above hypothesis, as shown in fig. \ref{fig:6}a. The plot shows that at room temperature as pressure around the sample is lowered to - 0.1 MPa from the ambient, in steps of 0.02 MPa, the sample resistance decreases by $\sim$ 2.5\% and then recovers once the chamber is vented to atmospheric pressure. Further, the sample resistance increases by $\sim$ 1\% if the chamber is pressurized to + 0.1 MPa above ambient, both being reversible changes. The later effect was reproduced with air, \ch{H2}, Ar, \ch{N2} and \ch{O2} negating any chemical origin to the change in resistance. 

Fig. \ref{fig:6}b shows the time variation in resistance as the surrounding pressure is decreased, which showed an initial spike before decreasing to a saturated value. The opposite effect was observed when the sample was pressurized above the ambient, where resistance increase with each increment in pressure is presaged by a resistance dip, as shown in fig. \ref{fig:6}c. 
These observations are explained as effect of  force on a porous system, as visualized in fig. \ref{fig:6}d. Upon evacuation the porous sample initially expands due to the excess internal pressure from the trapped gases, and later shrinks upon removal of the same. Thus initially the sample conductivity decreases and later increases due to relaxation and compression of  the Ni NPs together. The opposite mechanism is envisaged to comprehend the initial dip and later increase in resistance under excess pressurization. Slope of the resistance $vs.$ pressure curve (SI fig. S22) yields the pressure sensitivity of resistance to be $~$  0.1 kPa$^{-1}$. Further details of the experiments with variable pressure is available in SI section S6. The IMT is also perturbed by hydrostatic pressure as seen in the $RT$ plots in SI fig. S13. The figure plots T dependence of the sample resistance, R$_{1423}$ in C4P configuration with the sample in 10$^{-4}$ mbar vacuum and under 100 kPa (1 atm) pressure of \ch{N2}. Sample resistance is lower in vacuum than under 1 atm  pressure, in line with the prior discussion and both curves show the IMT transition though the $T_c$ is lower under pressure than in vacuum.  

To summarize, we have investigated anomalous electrical properties of a nanostructured hybrid material composed of a \ch{mTiO2} fibrous dendritic network decorated with polydispersed Ni nanoparticles. 
The highly porous semiconducting matrix displays a multiply-connected topology that induce anisotropic electrical transport in a polycrystalline bulk system. 
Four probe transport measurements in different contact configurations evidence the anisotropy with the high temperature state appearing as insulating or metallic dependent on the direction of measurement. 
The insulating phase undergoes transition to a low temperature metallic phase with T$_c \sim$ 100 K which shown to be likely associated with the unique sample morphology and its negative thermal expansion coefficient in the low temperature regime.
We also investigated absolute negative resistance observed below 100 K which is shown to arise from the highly anisotropic conductivity tensor. A combination of electrical circuit model and finite element analysis is used to simulate the carrier transport that result in voltage - current plots with negative slope. The simulations help visualize the current pathways and potential distribution that lead to the measurement of negative resistance. 
The analysis of anisotropic transport and negative resistance in this complex system challenges the traditional comprehension of four-probe transport measurements. It is anticipated that the ensuing modelling and discussions will be beneficial in analysing anomalous transport properties observed in other systems, especially hybrid materials.

\section{Acknowledgement}
Authors acknowledge Dr. Arijit Kayal (IISER TVM) for help in conducting various experiments. Authors acknowledge DST, Govt. of India (DST/INT/POL/P-44/2020) and NAWA Bilateral exchange of scientists (PPN/BIN/2019/1/00111) for financial support and ISTEM, Govt. of India for access to COMSOL Multiphysics. HG acknowledges PhD fellowship from DST INSPIRE, Govt. of India. KB, KK and DAP thank ENSEMBLE$^3$ Project (MAB/2020/14) which is carried out within the International Research Agendas Programme (IRAP) of the Foundation for Polish Science co-financed by the European Union under the European Regional Development Fund and Teaming for Excellence Horizon 2020 programme of the European Commission (GA No. 857543).

\section{Author Contributions}
HG, JM conceptualized the investigation, KB, KK, DAP synthesized the samples. HG conducted the experiments performed analysis; SC conducted the simulations. All authors contributed to writing the manuscript.  

\section{Conflicts of interest}
There are no conflicts to declare.

%% file: SI.tex
\section*{\begin{center}
		Supporting Information (SI)
\end{center}}
	
\section{Synthesis}
\begin{figure}[H]
	\begin{center}
		\includegraphics[width=0.67\textwidth]{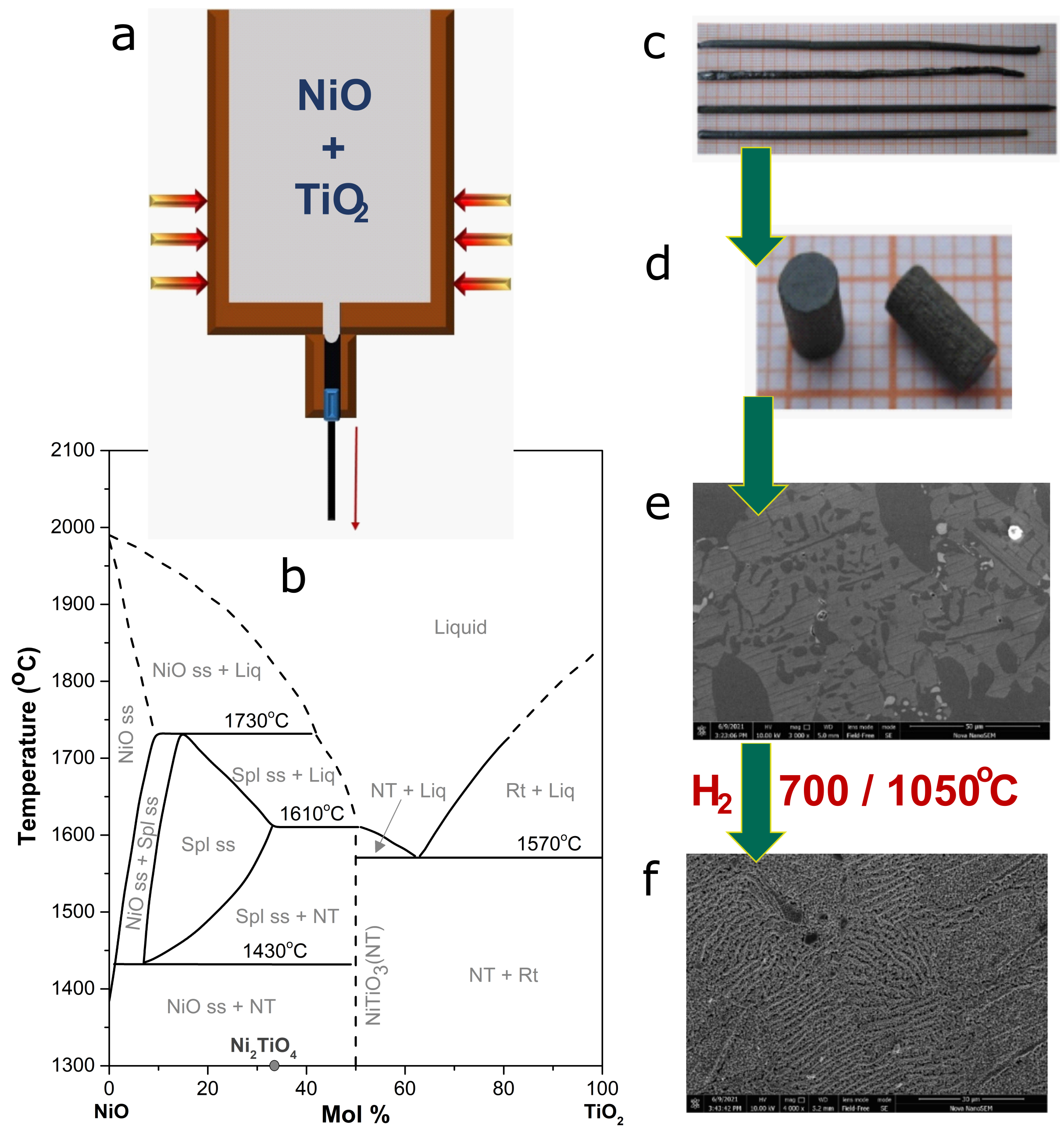}
		\caption{Scheme of Synthesis (a) $\mu$-PD method (b)Phase-diagram of NiO-\ch{TiO2}\cite{muan1992equilibrium} (c) optical image of synthesized eutectic rod (d) perpendicularly cut piece. SEM images of (e) polished eutectic surface (NTT) and (f) $1050^oC$ H$_2$ annealed sample(NTT1050).}
		\label{fig:S1}
	\end{center}
\end{figure}

\ch{NiTiO3}-\ch{TiO2} rods were grown by the micro-pulling-down method (µ-PD)\cite{fukuda2004fiber,yoon1994crystal,kolodziejak2016synthesis} in a nitrogen atmosphere following the phase diagram as in figure\ref{fig:S1}(b)\cite{muan1992equilibrium}. High-purity \ch{TiO2} (99.995\%) and NiO (99.999\%) powders (Alfa Aesar) were used as the starting materials. Materials were mixed (with over-eutectic composition of 69 mol\% \ch{TiO2} and 31 mol\% NiO) with pure 2-propanol in an alumina mortar. 
The obtained slurry was then annealed at 90$^oC$ for 2 h to remove potential volatile impurities. Then the obtained mixture was put in a Nabertherm N20/H oven at 1100$^oC$ for 10 h in $N_2$ or $Ar/CO_2$ atmosphere. The final product was used for the growth of \ch{NiTiO3}-\ch{TiO2} rods in the $\mu$-PD Cyberstar apparatus. The liquid flows in laminar fashion through the capillary and solidifies outside of the crucible die. The solidification starts after touching the melt at the die bottom with a seed (previously prepared \ch{NiTiO3-TiO2}) and the same will be retracted slowly with constant pulling rates, which conditions uniformity of eutectic materials and it yields rod shaped samples. The applied pulling rate was 3 mm/min. After the growth, composite rods were cut into small pieces (the thickness of the small disc$\sim$600 $\mu$m) perpendicular to the growth direction(figure\ref{fig:S1}(d)). 
Then the samples were polished with diamond lapping followed by annealing in Hydrogen($H_2$) atmosphere(100\% flow) at high temperatures (700, 1050$^oC$). The annealing step initiates and causes the reduction of nickel titanate\cite{arvanitidis2000intrinsic, morales2005hydrogen} to nickel and titanium dioxide (Eq\ref{eqn:synth}).
\begin{equation}\label{eqn:synth}
	NiTiO_3 + H_2 \rightarrow Ni + TiO_2 +H_2O \\   (884K \le T \le 1135K)
\end{equation}
Thus the nickel diffuses out and forms the NPs on TiO$_{2-\Delta}$ NW structure(fig.\ref{fig:S1}(f)). Hereafter samples \ch{NiTiO3-TiO2} is labelled NTT and the \ch{H2} annealed samples are named NTT700 and NTT1050, according to the annealing temperature.

\section{Characterization}

\subsection{Morphology and Elemental composition}
SEM and elemental mapping were done using Nova Nano SEM 450 fieldemission scanning electron microscope (SEM) coupled with an Apollo X energy dispersive X-ray analysis (EDS) system.
\begin{figure}[H]
	\begin{center}
		\includegraphics[width=0.98\textwidth]{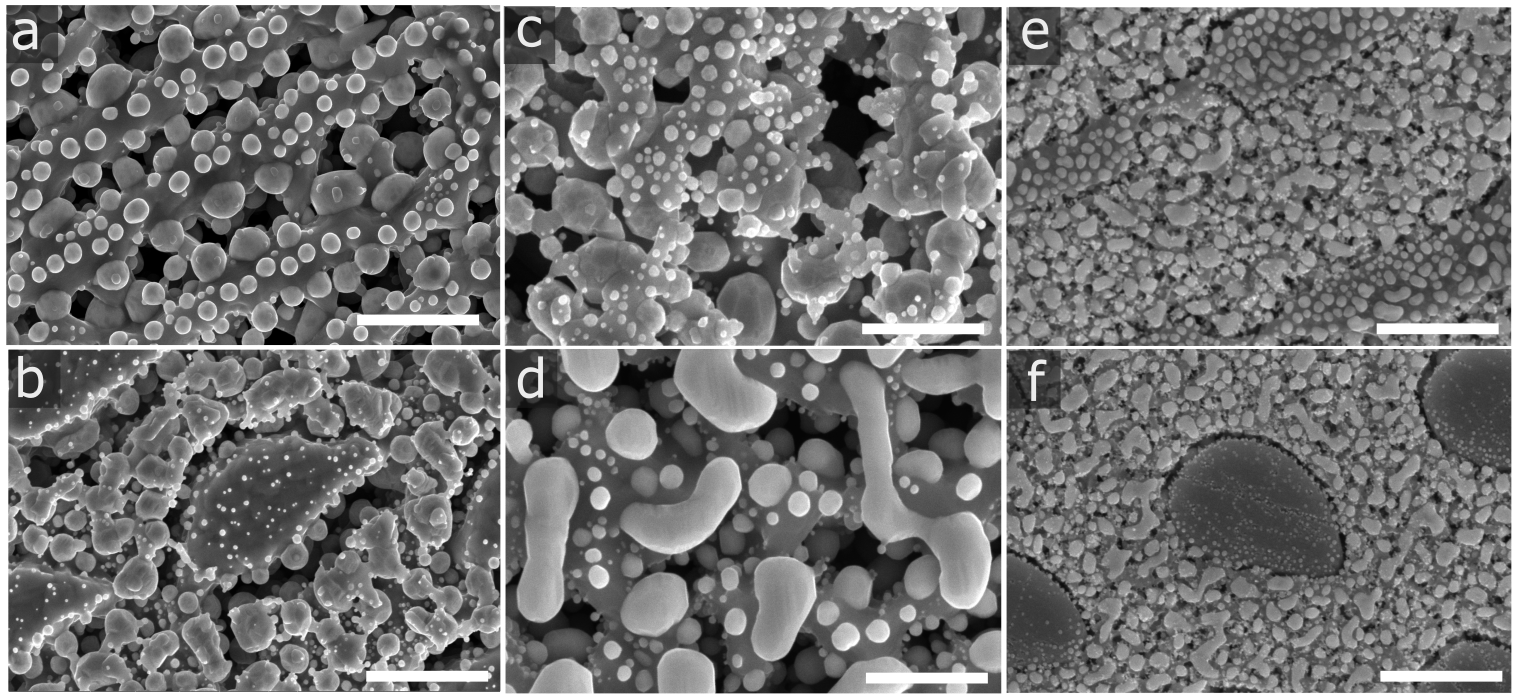}
		\caption{SEM images showing variation amongst \ch{H2} annealed NTT  samples (scale 2$\mu$m).}
		\label{fig:S1a}
	\end{center}
\end{figure}

\begin{table}[hpbt]
	\caption{Details of \ch{H2} annealed NTT samples shown in fig.\ref{fig:S1a}}
	\centering
	\begin{tabular}{|p{1.5cm} p{2.35cm} p{2cm} p{3.5cm}|}
		\hline
		Sample&Synthesis Environment&$\mu$-PD rate (mm/min)&\ch{H2} annealing: Temperature($^{\circ}$C) and Duration(min) \\ \hline
		\hline
		a&Ar/\ch{CO2}&3&1050 (20)\\
		b&\ch{N2}&0.15&1050 (20)\\
		c&Ar/\ch{CO2}&0.15&1050 (20)\\
		d&\ch{N2}&3&1050 (4)\\
		e&\ch{N2}&3&700 (20)\\
		f&Ar/\ch{CO2}&0.15&700 (20)\\ \hline	
	\end{tabular}
	\label{table5_1}
\end{table}

A conducting atomic force microscope (CAFM, MultiMode 8$^{TM}$ Bruker Inc.) was used to obtain the topography and current maps of the samples.
\begin{figure}[H]
	\begin{center}
		\includegraphics[width=0.75\textwidth]{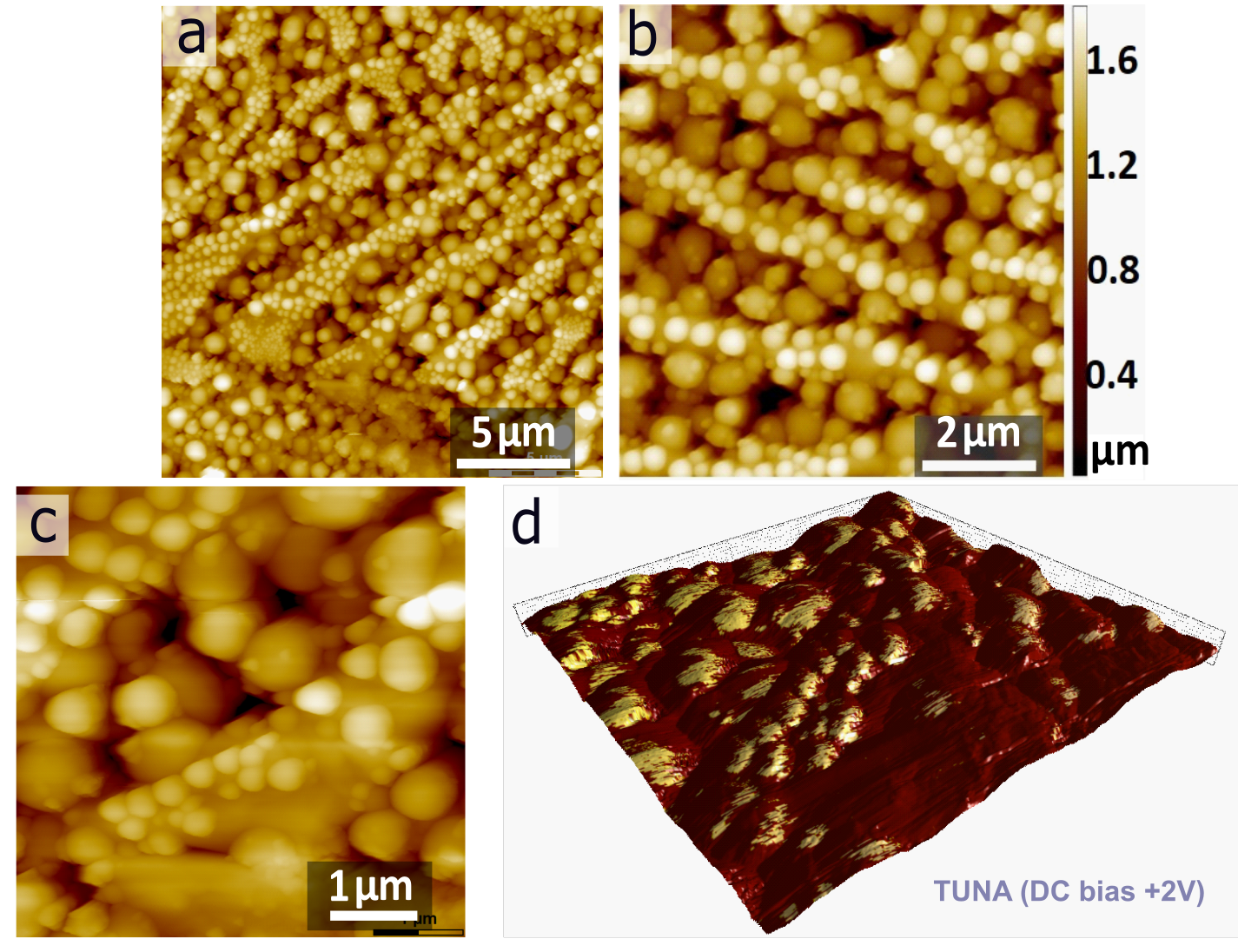}
		\caption{(a), (b) and (c) Topography and (d) overlayed TUNA (DC bias: 2V) of NTT1050 corresponding to (c).}
		\label{fig:S2c}
	\end{center}
\end{figure}

\begin{figure}[H]
	\begin{center}
		\includegraphics[width=0.95\textwidth]{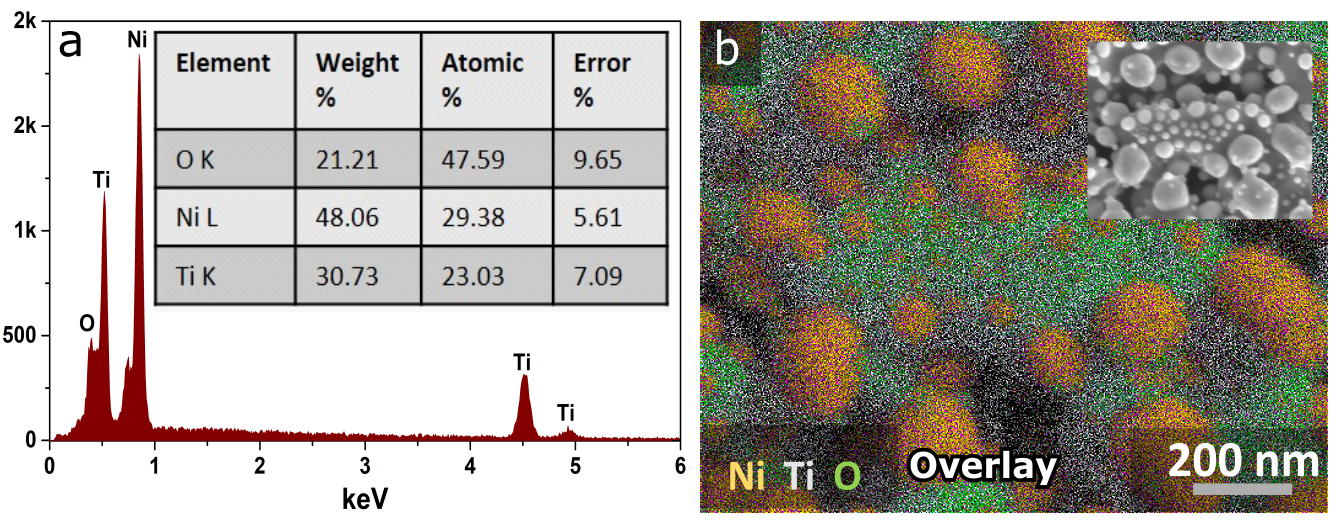}
		\caption{(a) EDS spectra(elemental percentages in the inset) and (b) Elemental mappping of NTT1050.}
		\label{fig:S2b}
	\end{center}
\end{figure}

\subsection{Structure and Stoichiometry}

XRD data was obtained from powdered samples using Rigaku SmartLab 3kW powder XRD (Cu K$\alpha$ – 1.5418\AA and operating Voltage 40kV) and analyzed using PDF4+2018 database.
\begin{figure}[H]
	\begin{center}
		\includegraphics[width=0.98\textwidth]{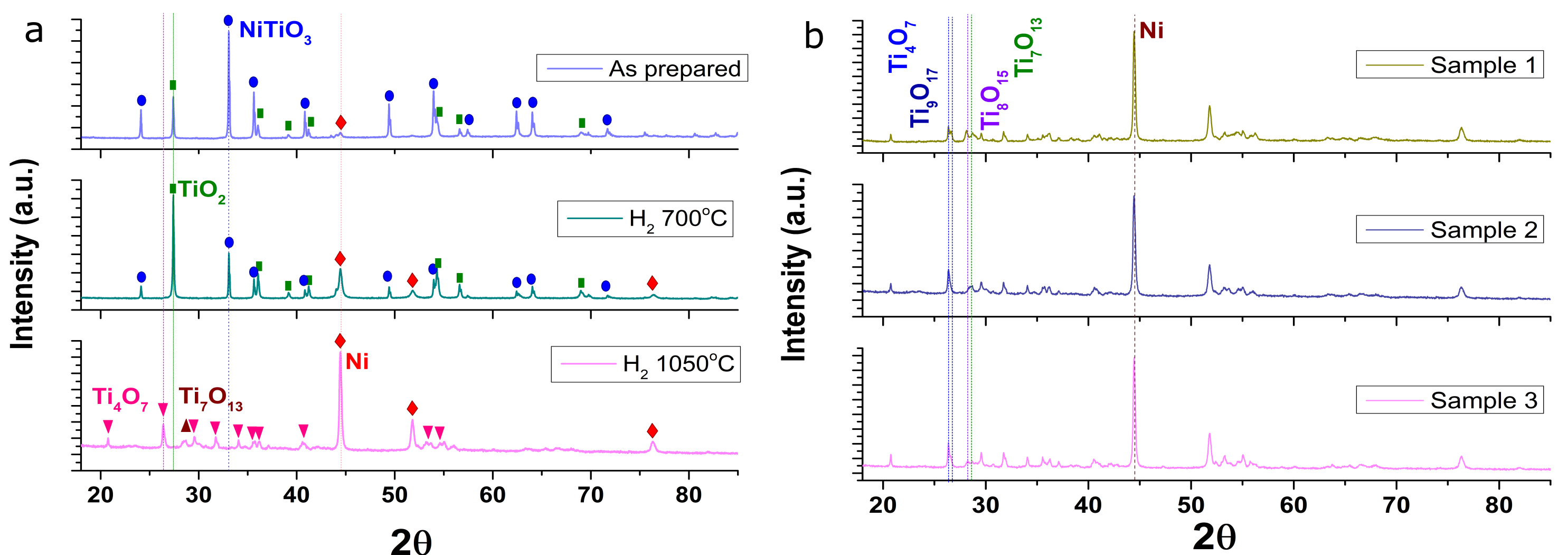}
		\caption{XRD data of (a) as prepared and annealed NTT samples and (b) different batches of NTT1050.}
		\label{fig:S2d}
	\end{center}
\end{figure}

Scienta Omicron ESCA+ was used to record XPS data with Mg K$\alpha$ followed by CasaXPS software for analysis.  The sample surface was cleaned by in-situ Argon etching prior to obtaining XPS data to avoid surface contaminants.
\begin{figure}[H]
	\begin{center}
		\includegraphics[width=0.85\textwidth]{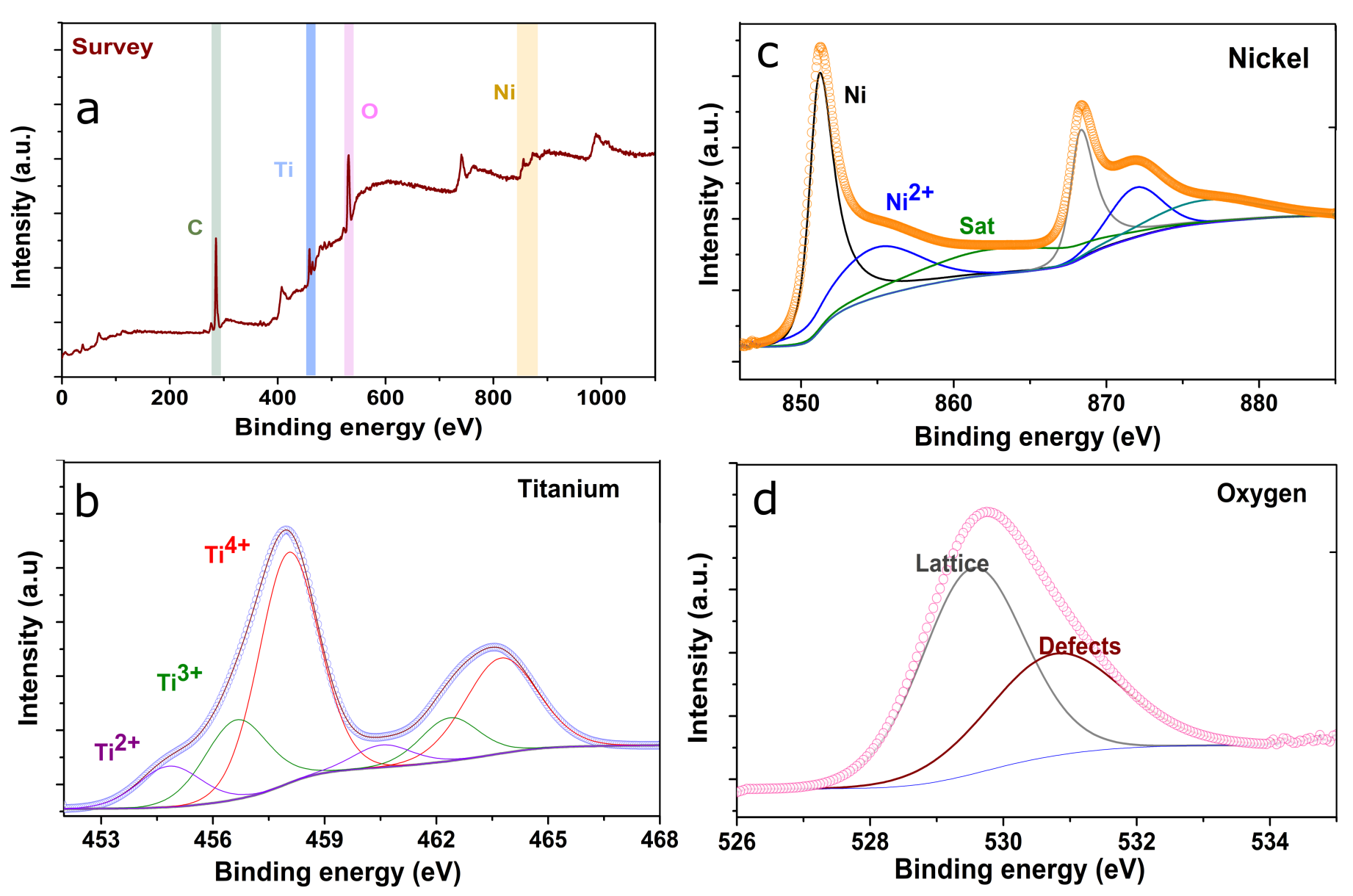}
		\caption{XPS(a)survey scan and high-resolution spectra of (b) Ti2p (c) Ni2p and (d) O1s core levels of NTT1050.}
		\label{fig:S2e}
	\end{center}
\end{figure}

\subsection{Continuity of the morphological nature}
\begin{figure}[H]
	\begin{center}
		\includegraphics[width=0.85\textwidth]{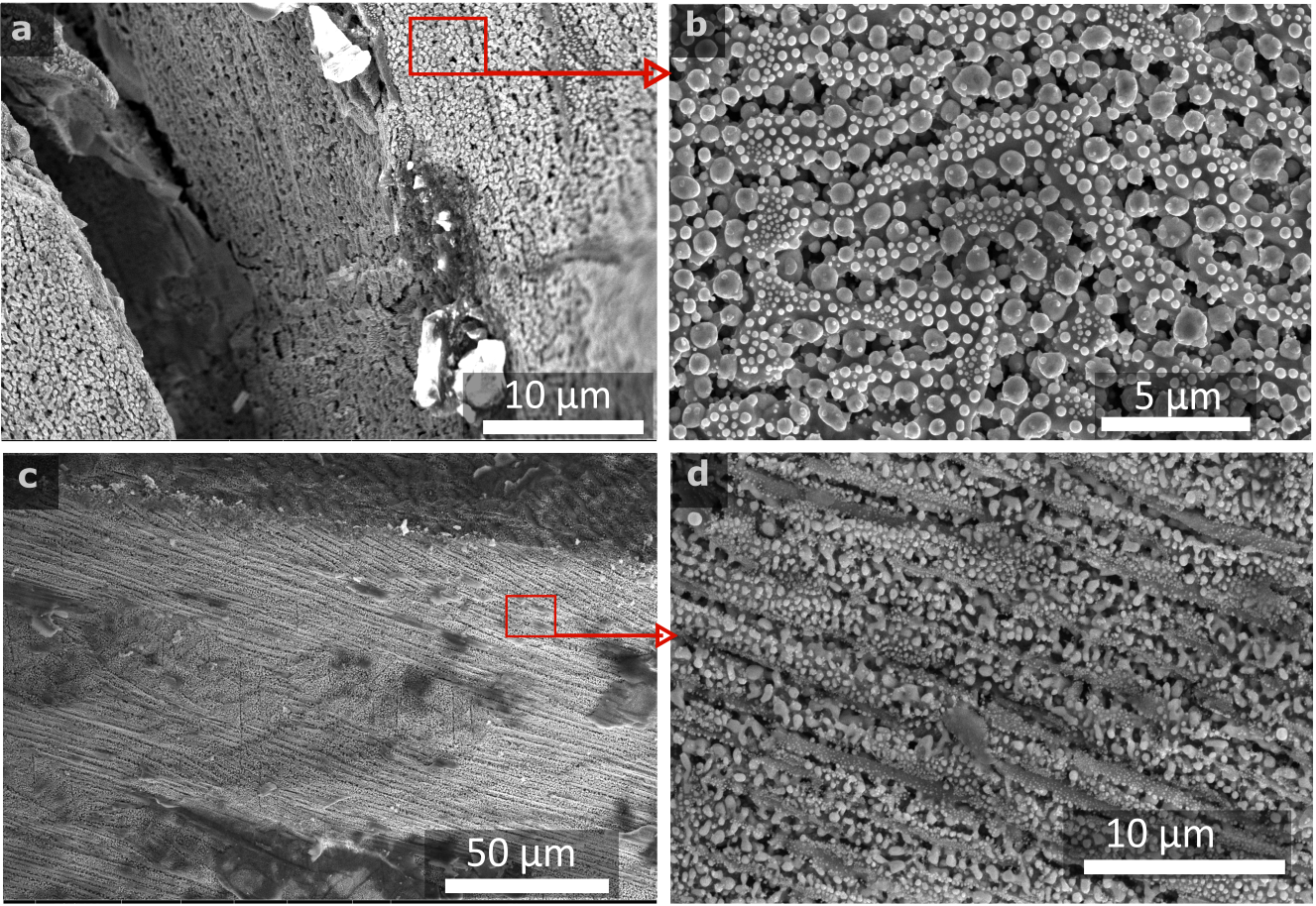}
		\caption{SEM images of (a) a crack on NTT1050 and (b) typical orientation of \ch{TiO2} decorated with Ni globules on the surface.(c) Broken side of NTT1050 with (d) zoomed region, showing the continuity of morphology along the depth of the sample.}
		\label{fig:S2a}
	\end{center}
\end{figure}

\section{Electrical measurements}
Room temperature electrical measurements were carried out in a custom made gas sensing chamber and the low temperature measurements inside a closed-cycle helium cryostat (ARS 4K). Electrical contacts were made on the sample with Silver(Ag) paint and Copper (Cu) wire and the sample is mounted on the cryostat holder with GE warnish ensuring proper thermal contact while doing temperature dependent measurements. Continuous monitoring of resistance (R) with temperature as well as Current-Voltage ($IV$) measurements with stabilized temperatures were conducted using Keithley 2400 Source meter.

\subsection{Low temperature}
\begin{figure}[H]
	\begin{center}
		\includegraphics[width=0.95\textwidth]{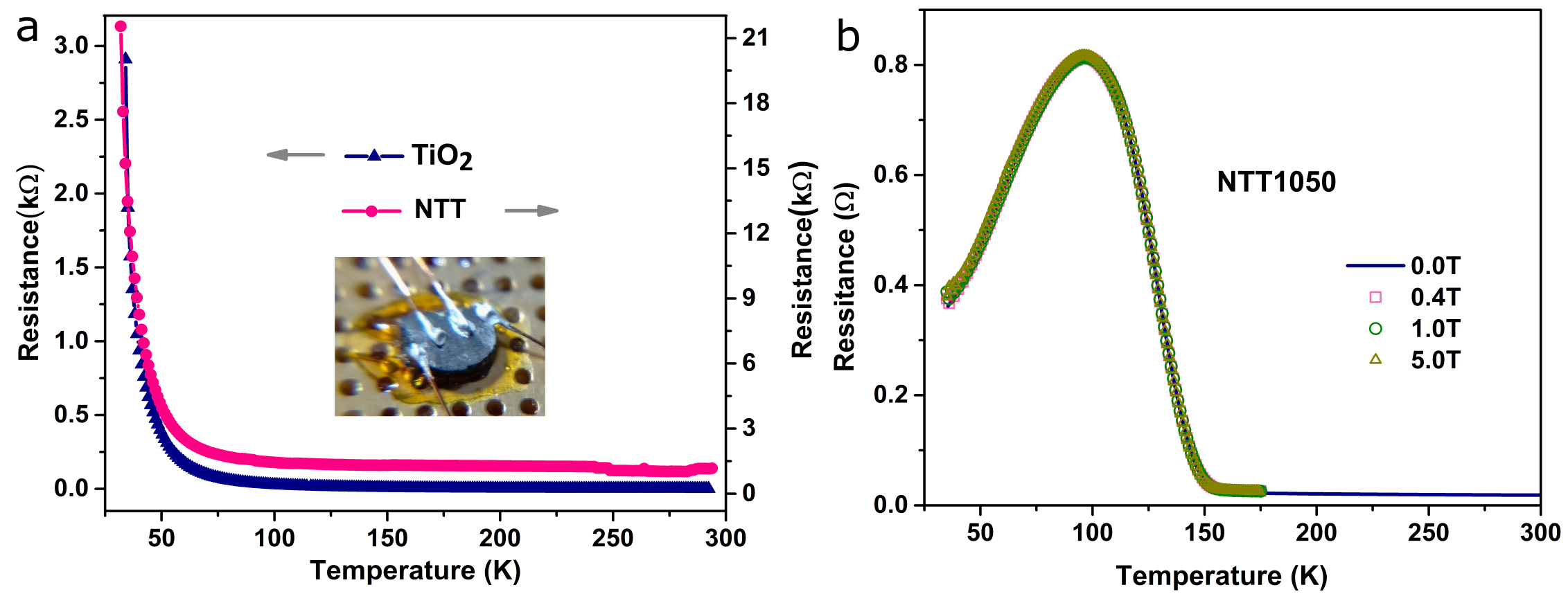}
		\caption{4P-$RT$ plots of (a) \ch{TiO2} and NTT. Inset shows the L4P contacts on NTT sample. (b) 4P-$RT$ of NTT1050 with and without applied magnetic field.}
		\label{fig:S3c}
	\end{center}
\end{figure}

\begin{figure}[H]
	\begin{center}
		\includegraphics[width=0.95\textwidth]{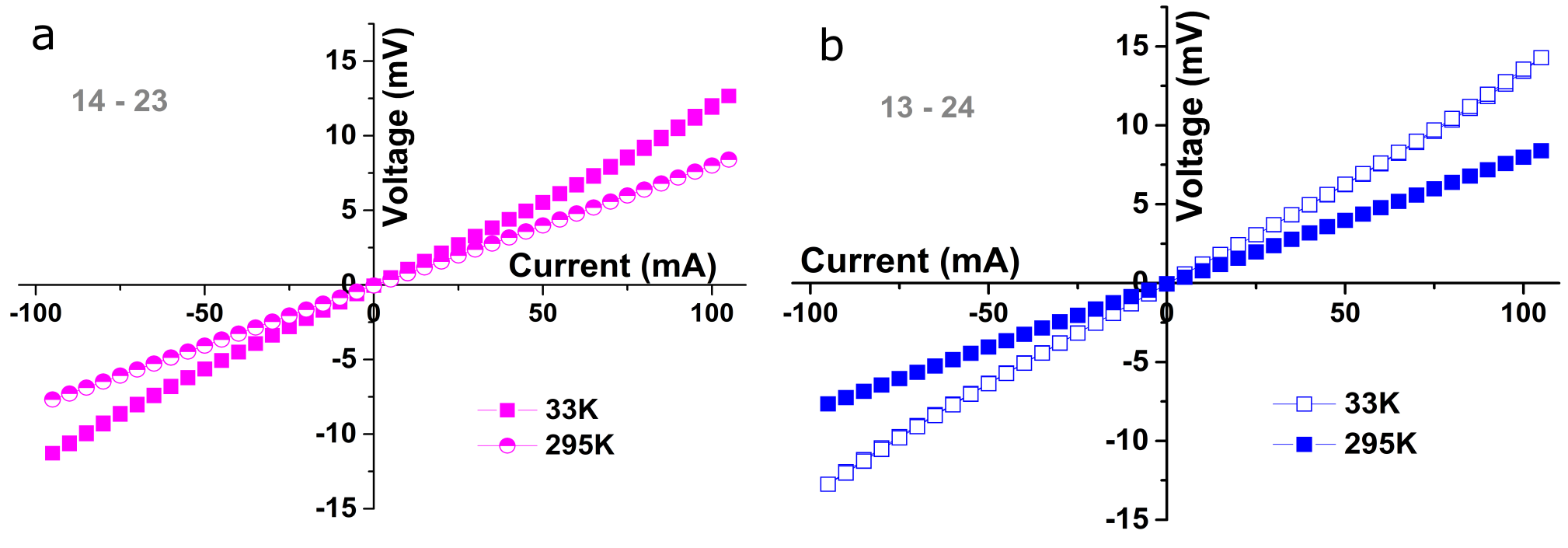}
		\caption{Room temperature and low temperature $VI$s of NTT1050 corresponding to (a) 14-23 and (b) 13-24 configurations.}
		\label{fig:S3b}
	\end{center}
\end{figure}

\begin{figure}[H]
	\begin{center}
		\includegraphics[width=\textwidth]{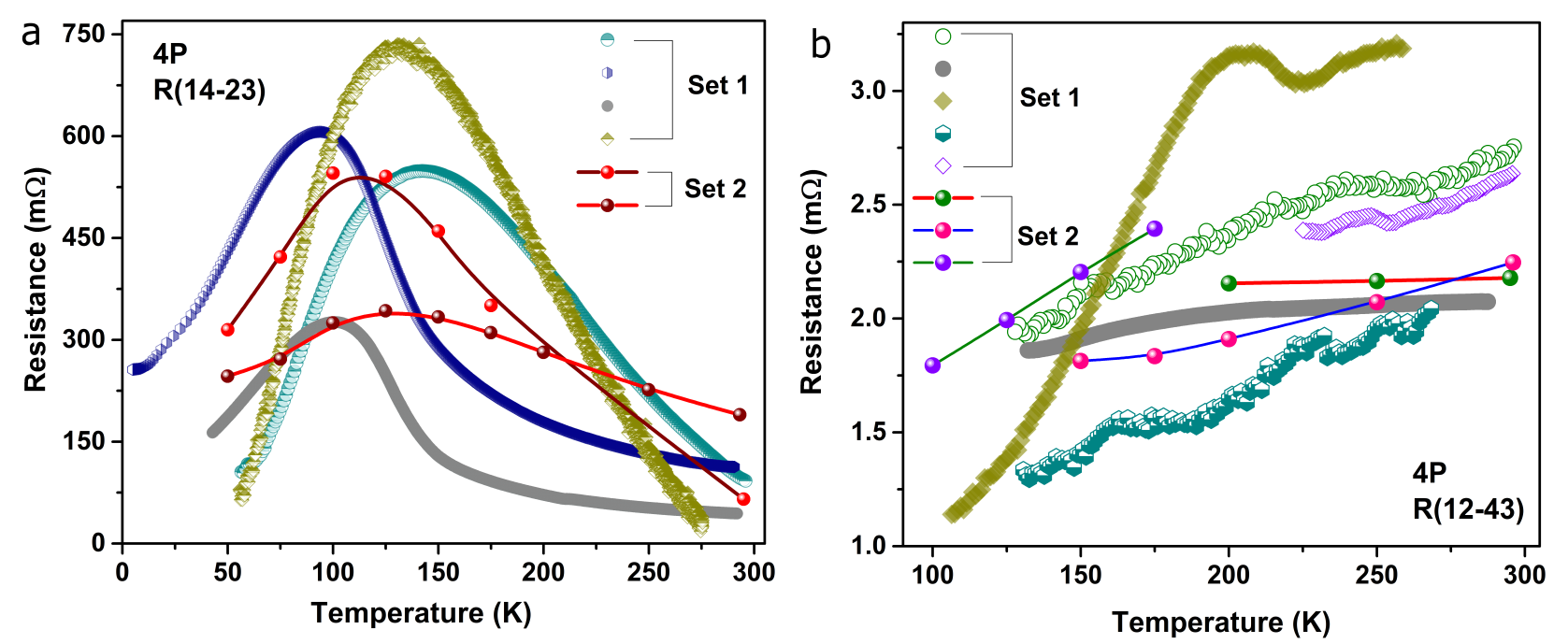}
		\caption{R$vs$T data of various NTT1050 samples with electrode configurations (a) 14-23 and (b) 12-43. Set 1 datas are obtained from $RT$ measurements, and set 2 datas are obtained from the T stabilised $VI$ (slope) measurements.}
		\label{fig:S3f}
	\end{center}
\end{figure}

\subsection{Analysis of contact configurations}
\begin{figure}[H]
	\begin{center}
		\includegraphics[width=0.65\textwidth]{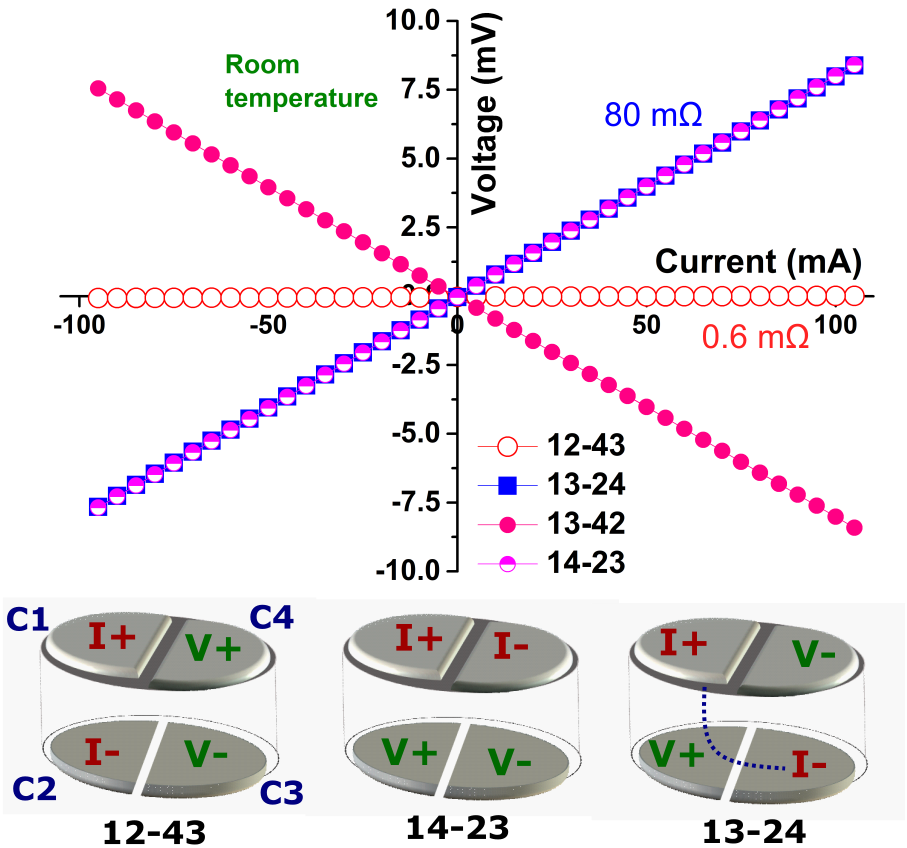}
		\caption{Room temperature $VI$s corresponding to different C4P configurations.}
		\label{fig:S3a}
	\end{center}
\end{figure}

\section{Structural changes with Temperature and Pressure}
	
	\subsection{Variation of $RT$ data with measurement atmosphere}
	Vacuum-free low-temperature measurements were done in \ch{N2} atmosphere using the temperature-dependent Raman set-up with modifications to include electrical connections.
	\begin{figure}[H]
		\begin{center}
			\includegraphics[width=0.4\textwidth]{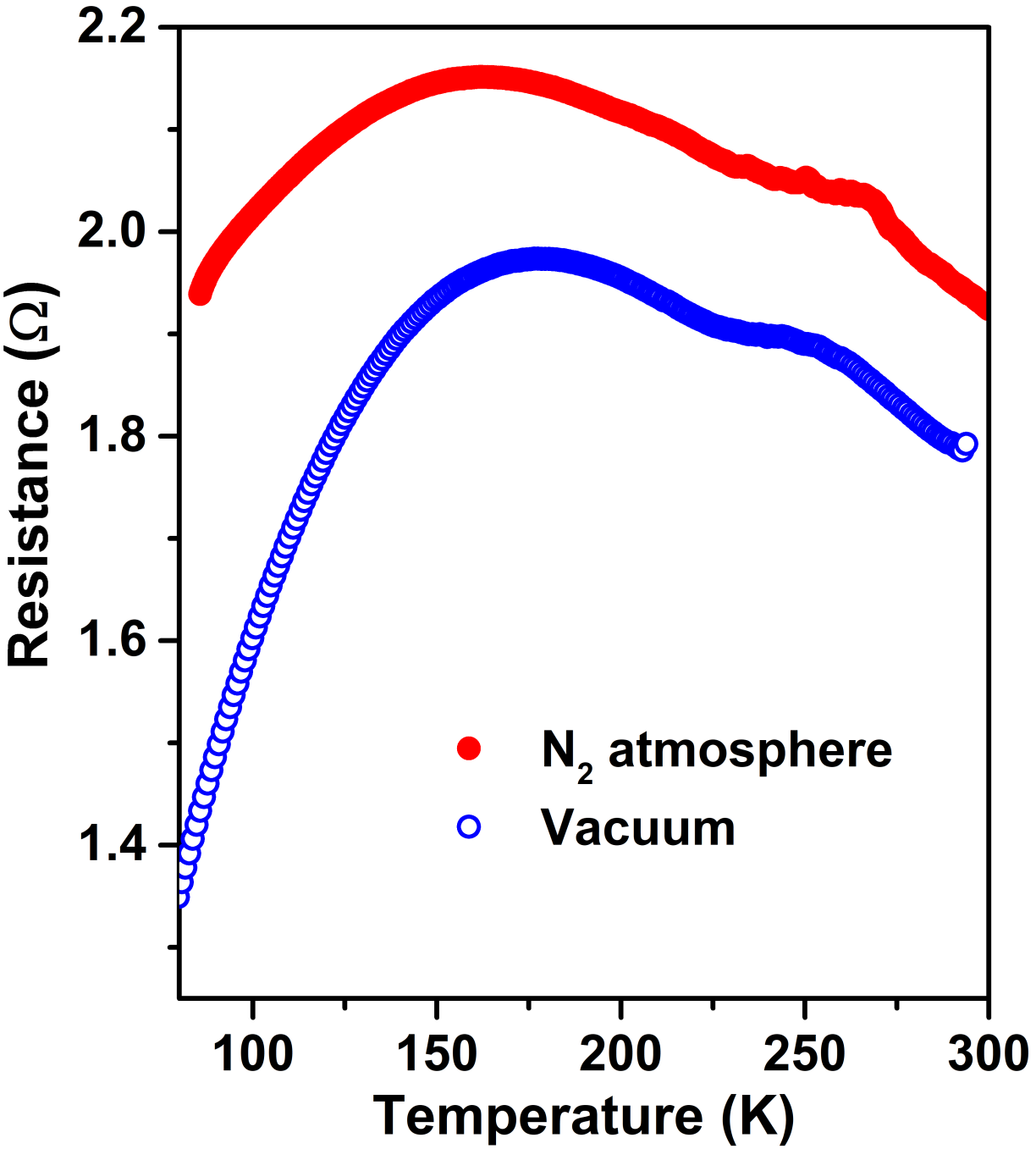}
			\caption{R$vs$T comparison: in Vacuum and \ch{N2} atmosphere in NTT1050.}
			\label{fig:S3d}
		\end{center}
	\end{figure}
	
	\subsection{Thermal expansion}
	The thermal expansion measurements were done with a K$\ddot{u}$chler’s dilatometer in the  physical property measurement system (PPMS-Dynacool) from Quantum Design Inc.
	\begin{figure}[H]
		\begin{center}
			\includegraphics[width=0.6\textwidth]{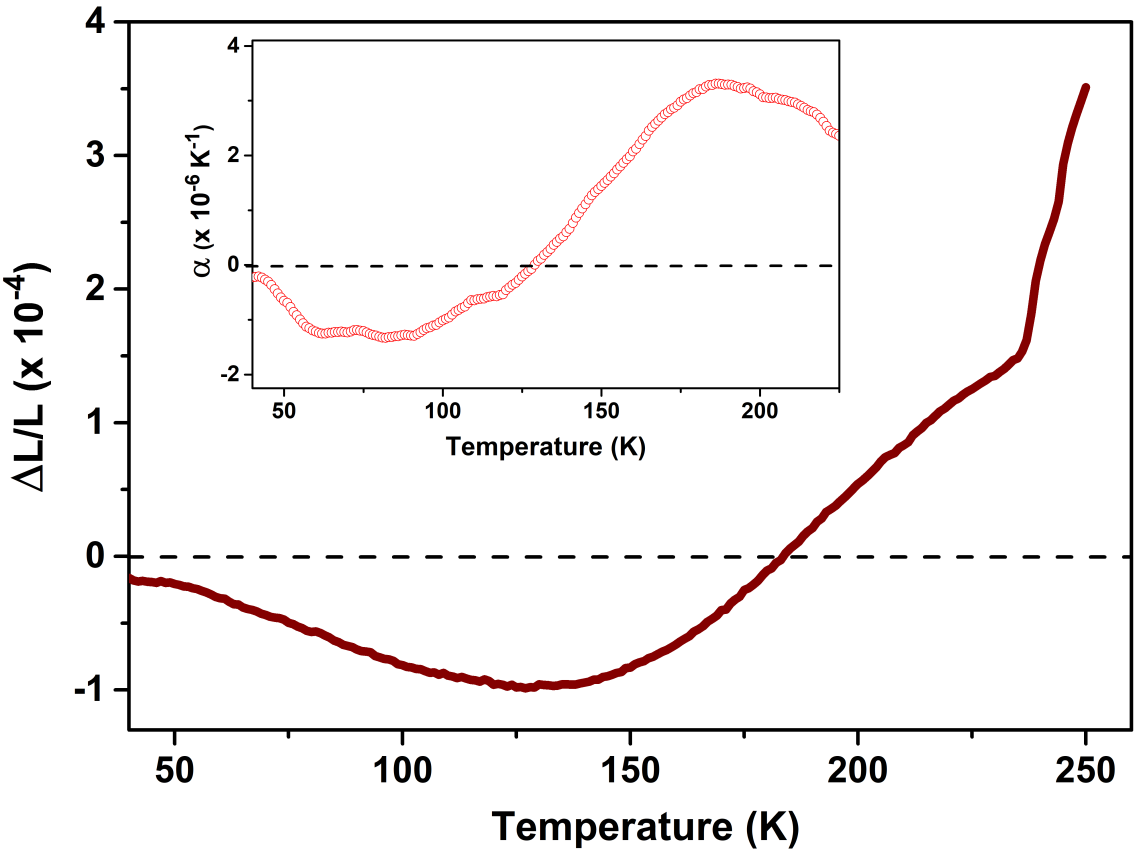}
			\caption{Calculated $\Delta$L/L as a function of temperature. Inset: linear thermal expansion coefficient at different temperatures.}
			\label{fig:S4a}
		\end{center}
	\end{figure}
	
	\section{Circuit modeling, Analytical calculations and FEM Simulations}
	
	\subsection{NI Multisim Circuit Simulator}
	
	NI Multisim software was used to simulate $VI$ curves where the R1, R2,and R3 values were chosen, guided by the values obtained from experimental $VI$ curves.
	
	\begin{table}[htbp]
		\caption{R values used fro circuit simulation.}
		\centering
		\begin{tabular}{|p{1.75cm}p{1.75cm}p{1.75cm}|}
			\hline
			&295K&35K\\ \hline
			\hline
			R1 &3.75$\Omega$&3.25$\Omega$\\
			R2&6.50$\Omega$&5.00$\Omega$\\
			R3&4.75$\Omega$&5.25$\Omega$\\ \hline
		\end{tabular}
		\label{table:T_1}
	\end{table}
	
	\subsection{Norton's equivalent Circuit}
	\begin{figure}[h!]
		\begin{center}
			\includegraphics[width=0.65\textwidth]{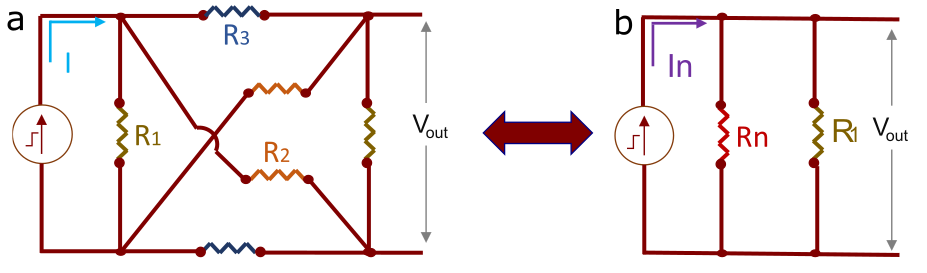}
			\caption{(a) The circuit model and (b) the corresponding Norton's equivalent circuit.}
			\label{fig:S4b1}
		\end{center}
	\end{figure}
	
	Analytical calculations were conducted to confirm the trend of 4P $RT$. A Norton’s equivalent circuit was formed for our circuit of interest (4 probe; 12-43), shown in fig.\ref{fig:S4b1}, and obtained the expressions for the effective R (R$_n$), effective current (I$_n$), current through the load (I$_l$) and the output voltage (V$_{out}$), as detailed below.
	
	Effective Norton Resistance ($R_n$)
	\begin{equation}
		R_n =ra +\dfrac{(rb+R2)(rc+R3)}{(rb+R2)+(rc+R3)}
	\end{equation}
	\begin{equation*}
		where, ra=\frac{R2R3}{R1+R2+R3}, rb=\frac{R1R3}{R1+R2+R3}, rc=\frac{R1R2}{R1+R2+R3}
	\end{equation*}
	effective current ($I_n$)
	\begin{equation}
		I_n =\dfrac{IR1(R2-R3)}{2R3(R1+R2)+R1(R2-R3)}
	\end{equation}
	where, I is the current sourced, 5mA\\
	current through the load ($I_l$)
	\begin{equation}
		I_l =\dfrac{I_nR_n}{(R1+R2))}
	\end{equation}
	output voltage (V$_{out}$)
	\begin{equation}
		V_{out}=I_lR1
	\end{equation}
	
\begin{figure}[H]
		\begin{center}
			\includegraphics[width=0.55\textwidth]{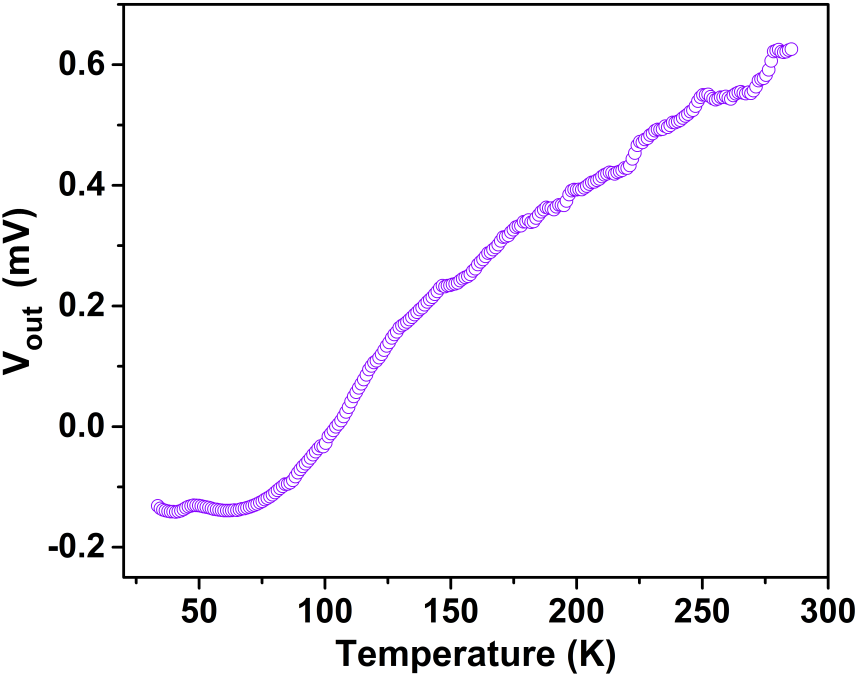}
			\caption{Analytical calculations derived variation of V$_{out}$ with Temperature, using the experimental 2-probe R values with 5mA current sourcing.}
			\label{fig:S4b2}
		\end{center}
	\end{figure}

	Quasi 4 probe $RT$ measurement values were fed into the equation of $I_n$, $R_n$, $I_l$ and V$_{out}$ in Matlab to get the V${_out}$ \textit{vs} T plot, in fig.\ref{fig:S4b2}.
	
	\subsection{Comsol simulations}
\begin{figure}[H]
		\begin{center}
			\includegraphics[width=0.85\textwidth]{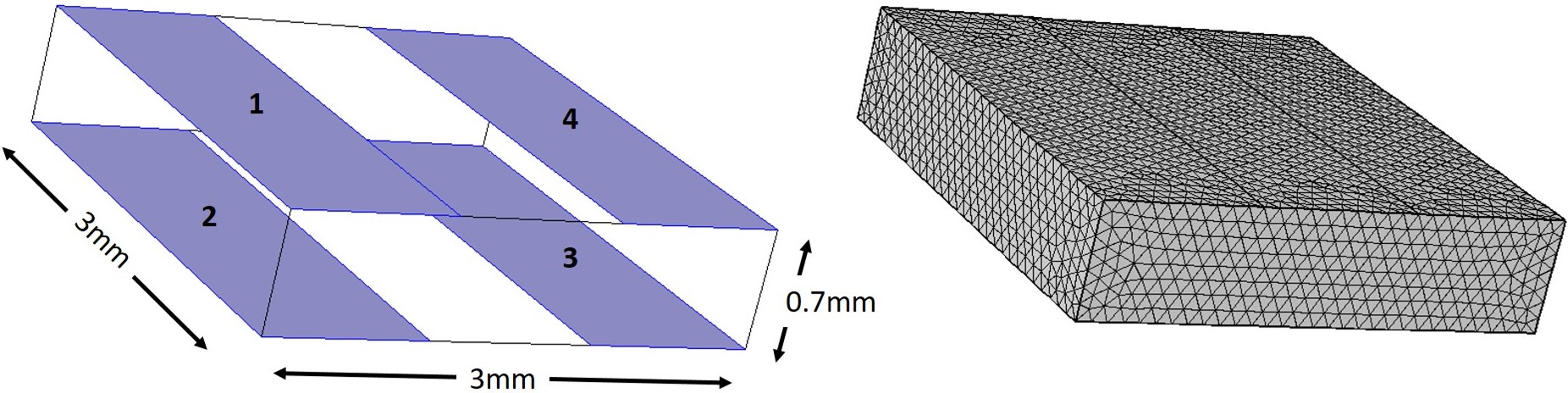}
			\caption{(a) Model geometry of the simulated sample displaying 4 contact through which current-voltage characteristics are acquired. (b) Meshing details of the modelled geometry.}
			\label{fig:S4c}
		\end{center}
	\end{figure}

	To simulate the current-voltage characteristics of a \ch{Ni-TiO2} eutectic sample using finite element modelling (FEM), a model was designed in in COMSOL 5.3(a). The model geometry is a cuboid having dimensions 3mm×3mm×0.7mm as shown in figure\ref{fig:S4c}a. Two contact terminals each were defined on top as well as on the bottom of the sample, each having dimensions 3mm×1mm labelled 1,2,3,4. The meshing density was kept at a minimum of 4.5 $\mu$m and a maximum of 0.1mm as shown in figure\ref{fig:S4c}b.
	The model is simulated using the Electric current module of COMSOL 5.3(a). Contact 1 is defined as a current terminal whereas contact 2 is defined as a ground terminal. The conductivity($\sigma$) of the model is given by 3x3 matrix $\sigma^{35}$ defined by using resistance values from experimental measurements (at 35K) as reference.  Hence the conductivity of the simulated sample with matrix $\sigma^{35}$ mimics the low temperature measurement at 35K.$\sigma_{zz}$ is kept significantly higher considering the electrode geometry.

	\begin{equation*}
		\sigma^{35}=\begin{bmatrix}
			1.78 & 3.90 & 0.10\\3.90 & 8.64 &2.50\\0.10 & 2.50 & 125
		\end{bmatrix}
	\end{equation*}
	Matrix $\sigma^{35}$ is a diagonalizable matrix with real positive diagonal terms by a transformation matrix $A$ to yield the diagonal matrix,$\sigma_p^{35}$. \\$A^T\sigma A$ = $\sigma_p$
	\begin{equation*}
		A=\begin{bmatrix}
			0.9104 & 0.4136 & 0.0015\\-0.4136 & 0.9102 &0.0215\\0.0075 & -0.0202 & 0.9998
		\end{bmatrix} ,
		\sigma_p^{35}=\begin{bmatrix}
			0.009 & 0.000 & 0.000\\0.000 & 10.36 &0.000\\0.000 & 0.000 & 125
		\end{bmatrix}
	\end{equation*}
	And the transformation matrix A satisfies the essential conditions for it to be a rotation matrix ($A^T=A^{-1}$ and $det(\abs{A})=1$).
	Similarly matrices for different temperatures are also constructed, by scaling the $\sigma$ values guided by the variation in measured R values at 75K, 125K and 285K respectively(table \ref{tab:1}). All these $\sigma$ matrices were also diagonalizable with real positive values. This could be interpreted as our material indeed have principal axis of conductivity($\sigma_p$), unlike an amorphous material. But the same is manifested as a $\sigma$ matrix with off-diagonal terms which could be understood to be originating from a rotation transformation.\\
	
\begin{figure}
		\begin{center}
			\includegraphics[width=0.85\textwidth]{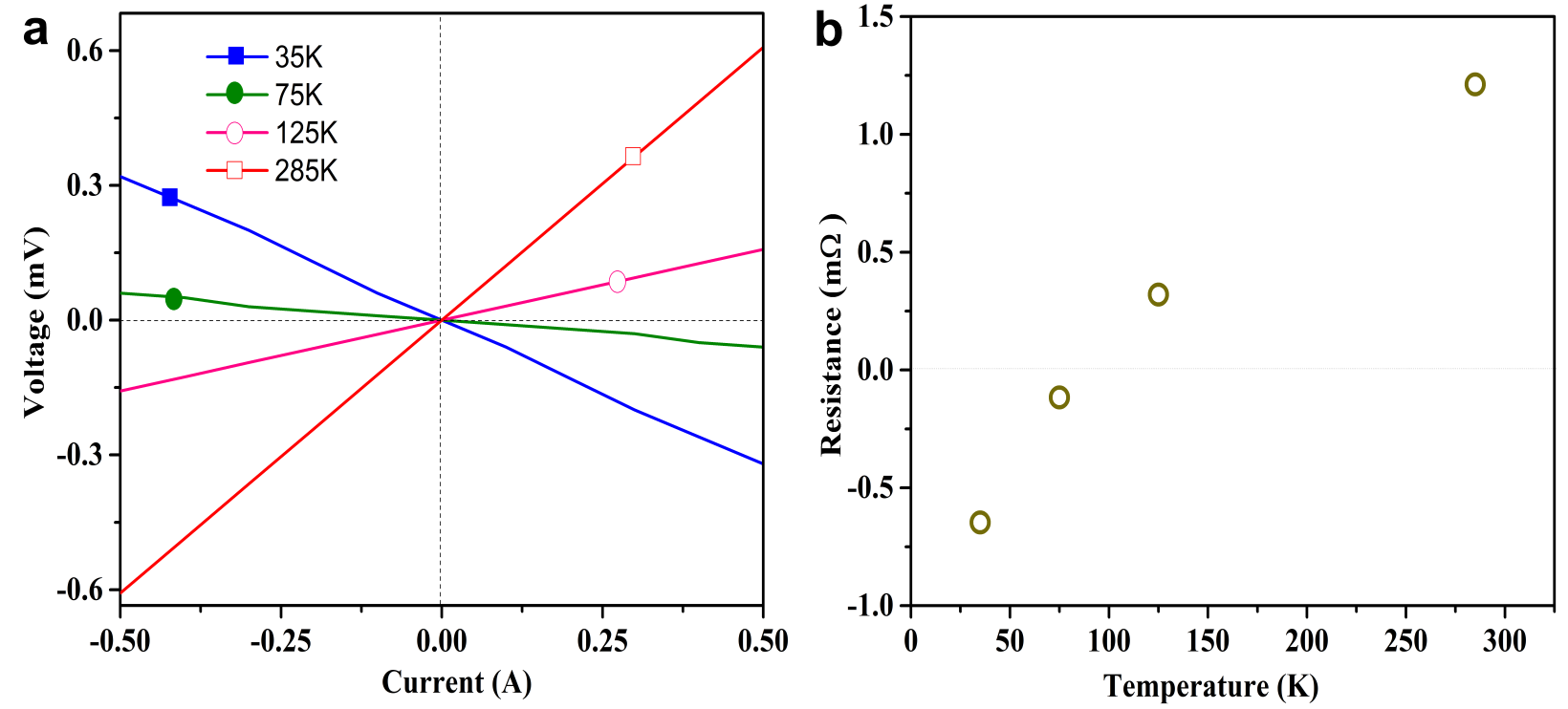}
			\caption{(a) $IV$ characteristics of the simulated system with 12-43 configuration and (b) the corresponding R (1/slope) at different temperatures.}
			\label{fig:S4d}
		\end{center}
	\end{figure}

\begin{figure}[H]
		\begin{center}
			\includegraphics[width=0.4\textwidth]{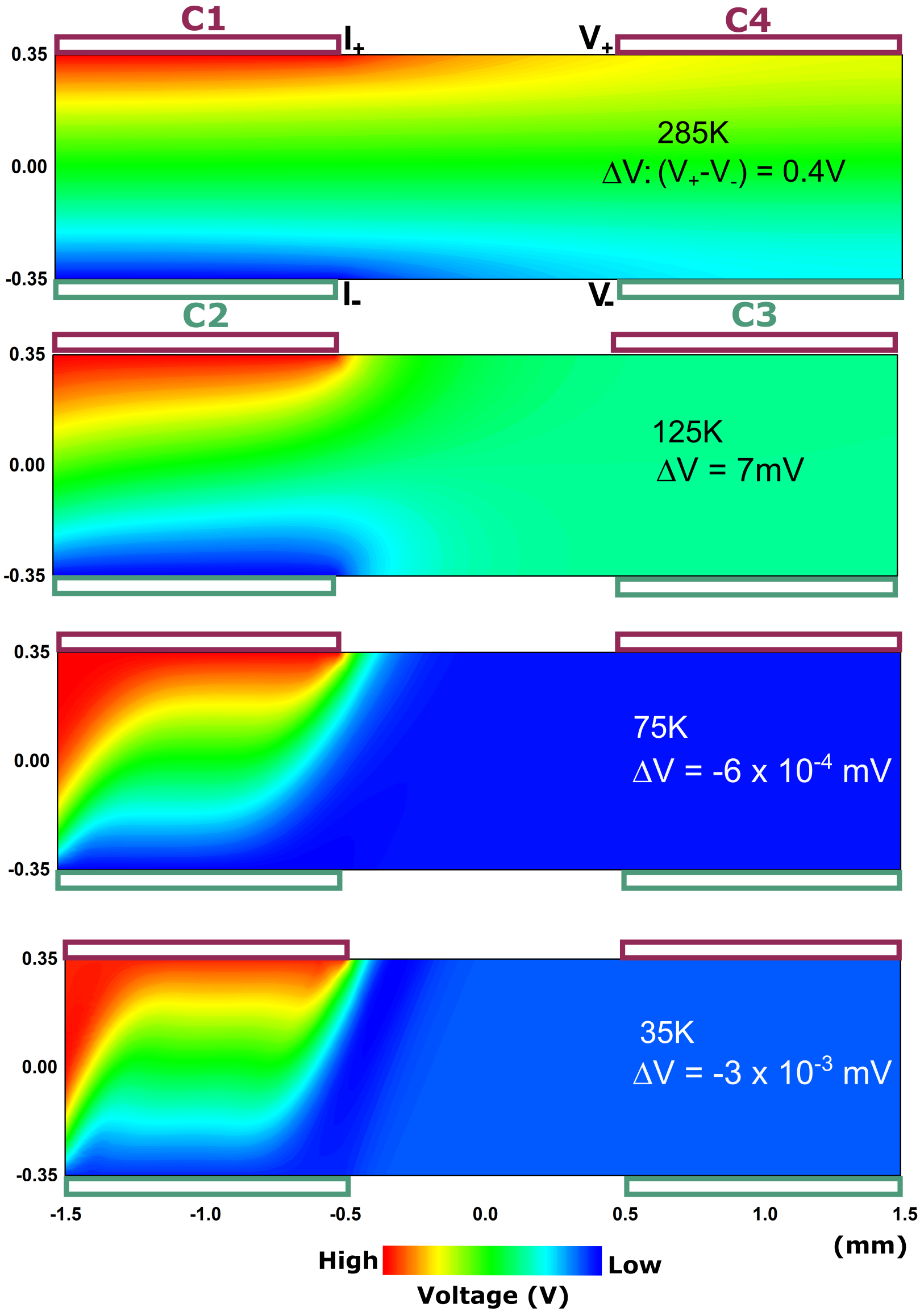}
			\caption{Surface potential plots with current sourcing at 1-2 contact terminal configuration and with corresponding $\sigma$ matrices of different temperatures.}
			\label{fig:S4e}
		\end{center}
	\end{figure}

	\begin{table}
		\caption{$\sigma, \sigma_p$ and $A$ matrices used for corresponding temperatures in the Comsol simulations}
		\begin{center}
			\begin{tabular}{ c c c c}
				\hline
				\hline
				& & & \\
				Temp. & \textbf{$\sigma$} & $A$ & \textbf{$\sigma_p$}\\
				& & & \\
				
				35K & $\begin{bmatrix}
					1.78 & 3.90 & 0.10\\3.90 & 8.64 &2.50\\0.10 & 2.50 & 125
				\end{bmatrix}$ &  $\begin{bmatrix}
					0.9104 & 0.4136 & 0.0015\\-0.4136 & 0.9102 &0.0215\\0.0075 & -0.0202 & 0.9998
				\end{bmatrix}$ &  $\begin{bmatrix}
					0.009 & 0.000 & 0.000\\0.000 & 10.36 &0.000\\0.000 & 0.000 & 125
				\end{bmatrix}$\\
				& & & \\
				75K & $\begin{bmatrix}
					1.17 & 1.71 & 0.02\\1.71 & 7.01 &0.43\\0.02 & 0.43 & 85
				\end{bmatrix}$ &  $\begin{bmatrix}
					0.9651 & 0.2619 & 0.0004\\-0.2619 & 0.9651 &0.0055\\0.0011 & -0.0054 & 1.0000
				\end{bmatrix}$ &  $\begin{bmatrix}
					0.71 & 0.000 & 0.000\\0.000 & 7.47 &0.000\\0.000 & 0.000 & 85
				\end{bmatrix}$\\
				& & & \\
				125K & $\begin{bmatrix}
					0.83 & 0.19 & 0.01\\0.19 & 0.57 &0.12\\0.01 & 0.12 & 60
				\end{bmatrix}$ &  $\begin{bmatrix}
					-0.4663 & 0.8846 & 0.0002\\0.8846 & 0.4663 &0.0020\\-0.0017 & -0.0011 & 1.0000
				\end{bmatrix}$ &  $\begin{bmatrix}
					0.47  & 0.000 & 0.000\\0.000 & 0.93 &0.000\\0.000 & 0.000 & 60
				\end{bmatrix}$\\
				& & & \\
				285K & $\begin{bmatrix}
					48 & 0.25 & 0.01\\0.25 & 0.61 &0.16\\0.01 & 0.16 & 3571
				\end{bmatrix}$ &  $\begin{bmatrix}
					-0.0052 & 1.0000 & 0.0000\\1.0000 & 0.0052 &0.000\\0.000 & 0.000 & 1.0000
				\end{bmatrix}$ &  $\begin{bmatrix}
					0.60  & 0.000 & 0.000\\0.000 & 48 &0.000\\0.000 & 0.000 & 3571
				\end{bmatrix}$\\
				& & & \\
				\hline
				\hline
				\label{tab:1}
			\end{tabular}
		\end{center}
	\end{table}
	$IV$ curves were simulated by defining the material conductivity by the $\sigma$ matrices at corresponding temperatures and are co-plotted in figure \ref{fig:S4d}.It is evident that simulation results corresponding to the conductivity matrix of temperature 35K and 75K has a negative slope i.e. an absolute negative resistance. With the increase in temperature, the corresponding $IV$ characteristics shift from negative to a positive slope similar to experimental results.  
	Figure\ref{fig:S4e} shows the 2D surface potential slice plot for 12-43 configuration at different $T$s. Clearly there is a potential flip causing the absolute negative resistance measurement. 
	
	\begin{figure}[H]
		\begin{center}
			\includegraphics[width=0.8\textwidth]{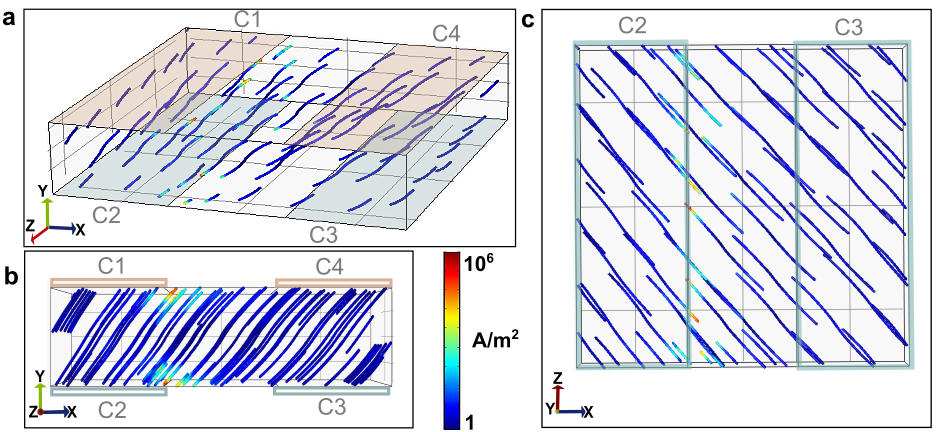}
			\caption{Streamline plots at 285K.}
			\label{fig:S4f}
		\end{center}
	\end{figure}

	\subsection{Validation of model with smaller and closer contacts}
	\begin{figure}[H]
		\begin{center}
			\includegraphics[width=0.6\textwidth]{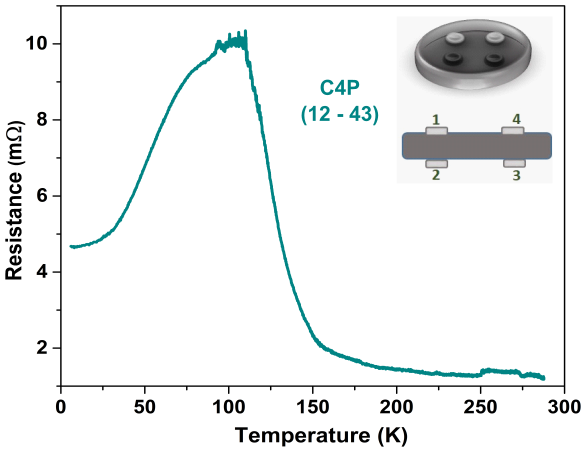}
			\caption{$RT$ of NTT1050 with smaller contacts.}
			\label{fig:S5a}
		\end{center}
	\end{figure}
	
	\section{Gas sensing set-up}
	
	\begin{figure}[H]
		\begin{center}
			\includegraphics[width=0.65\textwidth]{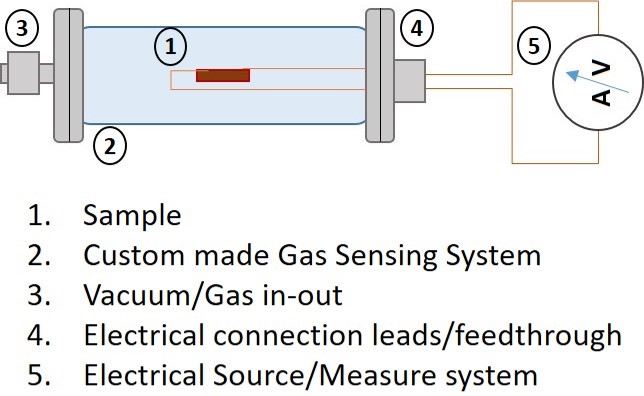}
			\caption{Gas sensing set-up.}
			\label{fig:S5}
		\end{center}
	\end{figure}
	
	\begin{figure}[H]
		\begin{center}
			\includegraphics[width=0.8\textwidth]{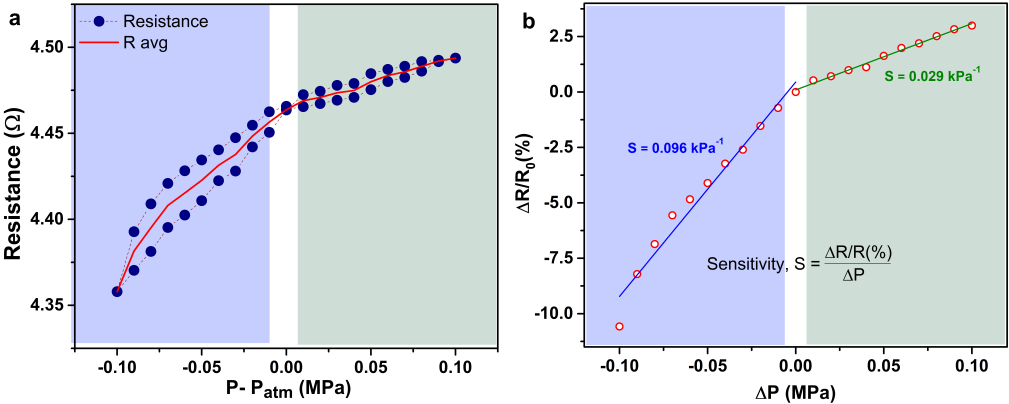}
			\caption{(a) Variation of R values with respect to the relative pressure inside the gas sensing chamber. (b) Sensitivity of the sample calculated from the $\Delta$R/R$_0$ $vs$ $\Delta$P plot, where $\Delta$R=R-R$_0$ and R$_0$ is the resistance at ambient conditions.}
			\label{fig:S5b}
		\end{center}
	\end{figure}